\documentclass[useAMS,usenatbib]{mn2e}

\usepackage{amsmath}
\usepackage{url}
\usepackage{amsfonts}
\usepackage{amsbsy}
\usepackage{graphicx}
\usepackage{subfigure}
\usepackage{verbatim}
\usepackage{amssymb}

\newcommand{\Rm}{\ensuremath{R_m}}

\renewcommand{\v}{\ensuremath{\mathbf{v}}}
\newcommand{\B}{\ensuremath{\mathbf{B}}}
\newcommand{\kk}{\ensuremath{\mathbf{k}}}

\renewcommand{\d}{\ensuremath{\partial}}

\newcommand{\ex}{\ensuremath{\mathbf{e}_{x}}}
\newcommand{\ey}{\ensuremath{\mathbf{e}_{y}}}
\newcommand{\ez}{\ensuremath{\mathbf{e}_{z}}}

\begin{document}

\title[MRI channel flows and their parasites]{MRI channel flows and their parasites}
\author[Henrik N. Latter, Pierre Lesaffre, and Steven A. Balbus  ]{Henrik N. Latter\thanks{E-mail: henrik.latter@lra.ens.fr},
Pierre Lesaffre\thanks{E-mail: pierre.lesaffre@lra.ens.fr},
 and Steven A. Balbus\thanks{E-mail: steven.balbus@lra.ens.fr}\\
Laboratoire de Radioastronomie,
 \'Ecole Normale Sup\'erieure,
 24 rue Lhomond, Paris 75005,  France}

\maketitle

\begin{abstract}
 Local simulations of the magnetorotational instability (MRI) in accretion
 disks can
exhibit recurrent coherent structures called channel flows. The
formation and destruction of these structures may play a role in the development
and saturation of MRI-induced turbulence, and consequently help us understand
 the time-dependent accretion behaviour of certain astrophysical objects.
Previous investigations have revealed that channel solutions are attacked by
various parasitic modes, foremost of which is an analogue of 
the Kelvin-Helmholtz instability. We revisit these instabilities and show how
they relate to the classical instabilities of plasma physics, the kink and
pinch modes. However, we argue that in most cases
  channels emerge from developed turbulence
and are eventually destroyed by turbulent mixing, not by the parasites.
 The exceptions are the clean isolated channels which appear
 in systems near criticality or which emerge from low amplitude initial conditions.
These structures  inevitably
 achieve large amplitudes and are only then destroyed, giving
 rise to eruptive behaviour.

\end{abstract}


\section{Introduction}

The initiator of magnetohydrodynamic turbulence in accretion disks, and
consequently of accretion itself, is most likely the magnetorotational
instability (MRI) (Balbus
and Hawley 1991, 1998). As its name suggests, the MRI marries
 differential rotation and magnetic tension which in combination draw energy from the
orbital motion and direct it into rapidly growing perturbations. 
 Numerical simulations show that the MRI excites
and sustains disordered flows that transport significant angular momentum
under a broad range of conditions (Hawley et
al.~1995, Hawley 2000). In addition, local simulations of the MRI
 exhibit recurring coherent
structures in which the flow vertically splits into two countermoving planar
streams (Sano 2007, Lesur and Longaretti 2007).
 These have been termed `channel flows' and coincide exactly with
unstable and axisymmetric linear MRI modes (Goodman and Xu 1994).
 It is to these structures, their stability and
role in MRI saturation, that this paper is devoted.

Goodman and Xu (1994) conducted the
 first analysis of the formation and breakup of channel flows.
 They revealed that these structures are
linearly unstable to two kinds of three-dimensional parasitic mode, the
fastest growing being a generalisation of the Kelvin-Helmholtz
instability. However, some simulations suggest that
 channel destruction may in many cases be a nonlinear process
 involving the interaction of a number of active MRI modes;
on the other hand, simulations which achieve strong magnetic fields
 report destabilisation by magnetic
reconnection (Hawley et al.~1995, Fleming et al.~2000, Lesaffre et
 al.~submitted).
The last point highlights the potential importance of plasma microphysics in the
development and saturation of the MRI, an issue that has greatly interested
researchers of late (Lesaffre and Balbus 2007, Fromang et al.~2007, Pessah et
al.~2007).

In this paper we investigate, with semi-analytic techniques and numerical simulations, 
the salient characteristics and dynamical role
of channel solutions in the development of the MRI.
 We first revisit the
study of Goodman and Xu (1994) and reinterpret their results, connecting
them to work in the field of magnetic jets, while also generalising the
analysis to include resistivity. Motivated by the appearance of thin
intense channels in our simulations, we also generalise the analysis to 
include compressibility and more `extreme' channel profiles.
These predictions are compared to compressible three-dimensional
simulations.
Lastly, we argue that the mechanism of channel breakdown in most cases
 is inherently
nonlinear, involving turbulent fluctuations, and is not a result of
linear parasitic modes.
We discuss briefly how this behaviour connects to the geometry of
the computational domain and the plasma beta parameter $\beta$.

\section{Goodman and Xu (1994) revisited}

The natural starting point of our discussion is the important study of Goodman
and Xu (1994), hereafter referred to as `GX'.
This work not only revealed that certain linear MRI modes are exact
 nonlinear solutions, it also detailed the various instabilities
to which they are subject. In this section we return to these results
and clarify a few issues pertaining to the behaviour of the secondary
parasitic instabilities. 

In addition, we generalise the results to non-ideal
MHD by introducing Ohmic resistivity. It is true in most astrophysical
situations, particularly those that involve fully ionised plasmas, that
  resistivities are exceedingly small (though in numerical simulations they
  must inconveniently take larger values). But
 a small
  resistivity can have important effects in a layer encompassing the magnetic null
  surfaces of the channel solution: within this layer 
 one class of parasitic mode functions similarly to the tearing mode and reconnects
 magnetic field.

  On the other
 hand, we neglect viscosity, mainly for simplicity, but also because we expect its
 impact to be minimal on the modes in which we are interested --- lowering
their  growth rates and slightly restricting instability to longer
 wavelengths. Viscosity will also alter the MRI channel structure itself, but
 this is only appreciable when the Reynolds number is extremely small, less
 than about 10 
(Pessah and Chan 2008), much smaller than in astrophysical systems and
 simulations. Recently, research has explored the
 connection between the magnetic Prandtl number and the saturation of MRI
 induced turbulence (Fromang et al.\ 2007, Lesur and Longaretti 2007).
 We do not believe, however, that
 this connection extends to parasitic modes as well, nor their dependence on
 viscosity. More likely, the magnetic Prandtl number dependence issues from
 the separation of the two dissipative scales, the viscous and resistive, and
 how this impacts on the turbulent cascade of energy 
(Balbus and Hawley 1998, Balbus and Henri 2008).

Before we plunge into the analysis we must first set up the problem, exhibit the
governing equations, and provide a little background to the MRI itself. This is
undertaken in the following section.

\subsection{Governing equations}

We study a disk of conducting fluid orbitting a massive compact object. Its motion is
described by the orbital frequency $\Omega(r)$ (a function of cylindrical radius $r$)
and it is threaded by vertical magnetic field $\B=B_0\,\ez$.
Initially we assume the imposed magnetic field is weak and the Alfv\'{e}n
speed much less than the sound speed. Typical fluid
velocities are also assumed much less than the sound speed because fluid motion is
(at first) triggered by
magnetic tension. As a
consequence, we adopt the equations of
incompressible nonideal MHD. 

The equations are framed in the 
geometry of the shearing sheet, which is a standard local approximation that
describes a
small `block' of disk centred at a radius $r_0$ moving on the circular
orbit $r_0$ prescribes.  The model provides an adequate
approximation to phenomena that vary on length-scales much less than
the large-scale properties of the disk.
 The block is represented in Cartesian coordinates with the $x$ and $y$
directions corresponding to the radial and azimuthal directions
respectively. To account for the differential rotation we add a Coriolis
force and impose a background linear shear, $\v= 2 A_0\, x\, \ey$, where 
$$ A_0 = \frac{r_0}{2}\left(\frac{d\Omega}{d r}\right)_0.$$
A Keplerian disk yields $A_0= -3/4$. Though we leave $A_0$ free in the
following, we nearly always let it take its Keplerian value for applications.

The governing equations are
\begin{align} 
&\frac{\d \v }{\d t} +\v\cdot\nabla\v =
-\frac{1}{\rho_0}\nabla\left(P+\frac{B^2}{8\pi}\right) +
\frac{\B\cdot\nabla\B}{4\pi\rho_0} \notag \\
&\hskip4cm - 2\mathbf{\Omega}\times\v -4A_0\Omega_0\, x\,\ex,
\label{u}\\ 
& \frac{\d\B}{\d t}= \nabla\times(\v\times\B) + \eta \nabla^2\B, \label{B}\\
&\nabla\cdot\B=0, \\
&\nabla\cdot\v=0 \label{inc}
\end{align}
where $\mathbf{\Omega}=\Omega_0 \ez$, and $\Omega_0=\Omega(r_0)$.
 The resistivity
is denoted by $\eta$ and $\rho_0$ is the homogeneous background mass density, 
a constant. To this set we must add appropriate boundary
 conditions. For convenience, the `0' subscripts on $A_0$ and
 $\Omega_0$ will be dropped.

By construction, the governing equations admit the equilibrium solution of 
linear shear and uniform vertical field: $\v= 2\, A\, x\, \ey$ and $\B= B_0\,
\ez$. But this configuration, as has been well established, is violently
unstable to the MRI for a sufficiently weak imposed $\B$ (Balbus and Hawley 1991). Small
perturbations 
of the form $e^{st+ i K z}$ grow rapidly and exhibit growth rates
of order an orbital period. The most vigorous mode grows at a rate $s=-A$.

The instability
gives rise to a vertical sequence of
planar countermoving fluid streams, or channels. What is remarkable is that this
configuration not only corresponds to the linear eigenfunction of the MRI 
but to a nonsteady but exact nonlinear solution to the 
set Eqs~\eqref{u}-\eqref{inc} (GX).
As a consequence, in our discussion on channels
 it will be as an exact nonlinear solution that we treat the MRI, not
 as a Fourier-decomposed linear mode. In the next section it is described
 mathematically.

\subsection{MRI channels}

 We adhere to the approach of GX,
differing only in the introduction of resistivity.
 Our ansatz for the MRI channel
solution is
\begin{align} \label{ch1}
& \v^{\text{ch}}= b\, e^{st}\,v_0\,\sin (Kz)\,(\ex\cos\theta+ \ey\sin\theta) \\
& \B^\text{ch} = b\,e^{st}\, B_0\,\cos (Kz)\,(\ex\cos\phi+ \ey\sin\phi) \label{ch2}
\end{align}
where $s$ is the growth rate of the mode, $K$ is its vertical wavenumber, $b$
is a dimensionless measure of the channel amplitude, and $v_0$ and $B_0$ are
dimensional constants. Note that there exist two orientation angles $\theta$
and $\phi$; if resistivity were absent we would only require one, because the
$\B$ field is then perpendicular to $\v$ (i.e.\ $\phi=\theta-\pi/2$).
 
These functions are superimposed on the background
linear shear and vertical field, so that 
$$ \v=\v^{\text{ch}}+2A\,x\,\ey, \qquad \B= \B^\text{ch}+B_0\,\ez, $$
and then substituted
into the $x$ and $y$ components of the governing equations \eqref{u} and
\eqref{B}. Because
$\v^\text{ch}\cdot\nabla\v^\text{ch}=\B^\text{ch}\cdot\nabla\B^\text{ch}=0$
and $\v^\text{ch}\cdot\nabla\B^\text{ch}=\B^\text{ch}\cdot\nabla\v^\text{ch}=0$
the nonlinear equations coincide with their linearised counterparts,
 and the solution is straightforward.
 It yields the dimensionless growth rate $s/\Omega$, the scaled flow speed
 $v_0 K/\Omega$, the scaled channel wavenumber $v_A K/\Omega$, and the orientation of the
magnetic field $\phi$, as functions of the
dimensionless inputs $A/\Omega$, $\theta$, and
$\Rm$, where the Alfv\'{e}n velocity and magnetic
Reynolds number are
 defined through
$$ v_A\equiv \frac{B_0}{\sqrt{4\pi\rho_0}},\qquad \qquad \Rm\equiv \frac{\Omega}{\eta K^2}.$$
 After a little algebra one finds
\begin{align} \label{sGX}
s &= -A\,\sin 2\theta - \frac{\Omega}{\Rm}\,\left(1+ \frac{A}{\Omega}\,\cos^2\theta \right),\\
\left(v_A K\right)^2 &= -4\,\Omega\,A\,\sin^2\theta\left(1+
  \frac{A}{\Omega}\,\cos^2\theta\right)\left(1+ \tfrac{1}{2}\Rm^{-1}\,\cot\theta\right)^2,
\label{vaGX} \\
v_0 &= -2\frac{A}{K}\,\sin^2\theta  \left(1+
  \tfrac{1}{2}\Rm^{-1}\cot\theta\right)^2\,\left(1+\tfrac{1}{4}\Rm^{-2}\right)^{-1/2}, \label{v0GX} \\
\phi &= \left(\theta- \frac{\pi}{2}\right) + \cot^{-1}\left(2\Rm\right), \label{opangle}
\end{align}
 In the limit of $\Rm\to\infty$ the GX ideal
MHD expressions are recovered, cf.\ their Eqs~(5)-(7). A large $\Rm$ introduces
a small correction of $1/\Rm$ to these expressions. 

The dispersion relation is a quartic in
 $s$ and can be computed by eliminating $\theta$ from equations \eqref{sGX}
and \eqref{vaGX}. It is
$$ (s\tilde{s} + v_A^2 K^2)^2 + 4\Omega(\Omega+A)\tilde{s}^2
 +  4A \Omega v_A^2 K^2 =0,$$
with $\tilde{s}= s + \eta K^2$. This agrees with a number of examples in the
 literature; for instance, see Sano and Miyama (1999), Fleming et al.~ (2000),
 Lesur and Longaretti (2007), or Pessah and Chan (2008). The most vigorously growing channel takes
 an orientation of 
$$ \theta_\text{max}= \frac{\pi}{4}+ \frac{1}{2}\,\cot^{-1}(2\Rm) $$
and exhibits the growth rate
$$ s_\text{max}= - A \sqrt{1+ \tfrac{1}{4}\Rm^{-2}} -
\frac{1}{2\Rm}(2\Omega+A).$$
At leading order in large $\Rm$ we recover $\theta=\pi/4$ and
$s_\text{max}=-A$, as expected.

Equation \eqref{vaGX} tells us that $(v_A K)^2$ is a decreasing function of
$\theta$. Therefore, longer wavelength MRI modes possess flows that are more
radial and magnetic fields that are more toroidal. In fact, $\theta\to
0$ as $K$
approaches 0 (the marginal stability limit of very long modes).
 In contrast, $\theta$ approaches a value near
$\pi/2$ as $K$ approaches $K_c$ (the marginal stability limit of
short modes).
 Thus short wavelength MRI modes near criticality possess flows nearly azimuthal
and magnetic fields nearly radial.

 The
Alfv\'{e}nic Mach number is defined by $M_A\equiv v_0/v_A$, and can be
computed from \eqref{vaGX} and \eqref{v0GX}.
We find that the strongest growing channels are characterised by fluid speeds
comparable and less than the Alfv\'{e}n velocity, $M_A \lesssim
1$. This is because the acceleration which drives the flow issues
from magnetic tension. Generally, sub-Alfv\'{e}nic flows are stable to 
shear instabilities because of the (relatively) large magnetic tension
(Chandrasekhar 1961). That being so, shear instabilities in MRI channel flows may
 orient themselves perpendicular to $\B^\text{ch}$ and escape this
constraint. Happily for them, $\B^\text{ch}$ and $\v^\text{ch}$ are nearly
perpendicular, and it is this fortuitous orientation that allows so much scope
for the parasitic instabilities that afflict channel solutions.

Finally, from Eq.~\eqref{opangle} it is clear that decreasing $\Rm$ also
decreases the opening angle between the magnetic and velocity fields. The
effect, however, is rather small until $\Rm\lesssim 10$. In most ionised plasma
the angle should be virtually 90 degrees; but in the interiors of 
protostellar disks it may be somewhat less.

\vskip0.3cm

Being a nonlinear solution, a channel mode will retain its structure and 
grow exponentially regardless of its amplitude. Unperturbed it will continue
to do so until the approximation of
incompressiblity fails. This will happen when
 the Alfv\'{e}n speed, and the speed of the
 fluid jets, approach the sound speed of the fluid itself. At this point in
 the evolution,
layers of magnetic field either side of a channel
 will give rise to a magnetic pressure `sink' sufficiently strong
 to squeeze mass towards the magnetic null surfaces at $Kz=(2n+1)\,\pi/2$ (for
 $n=0,1,\dots$).
 These altitudes are also where the fluid
 jets are maximal.
 As a consequence, the channel structure will
evolve from the sinusoidal profiles of \eqref{ch1} and \eqref{ch2}
towards a more `extreme' configuration. In fact, if undisturbed, we may
end up with a sequence of
 discontinuous
current sheets concentrating nearly all the mass and all the velocity upon
the magnetic null surfaces. Such sheets were briefly analysed by GX
 who rightly exhibited them as the opposite (compressible) limit to the
 incompressible MRI channels (see Fig.~13 later for a numerical approximation).

In any case, if an incompressible MRI channel is to reach the amplitudes at which
compressibility plays a role, the channel must grow undisrupted
 for a significant amount of time. One could be forgiven for thinking that 
this may not be
all that frequent an occurrence.
  On one hand, early in its
growth, an MRI channel may
interact with a different competing MRI mode, and this interaction could lead
to the early disruption of both.
 On the other hand, parasitic instabilities may emerge upon
the channel structure feeding off its energy and ultimately leading to its
demise.
 For the remainder of this section we examine the latter possibility;
in so doing we rediscuss the analysis of GX and
connect it to relevant work in the field of magnetohydrodynamical jets and
magnetic reconnection. Once this is finished we suggest that, in fact, the
large amplitude regime is more likely than one might think.

\subsection{Parasitic modes}

The linear stability of nonlinear 
channel solutions is complicated for a number of reasons, not least
 by the fact that the channels
are time-dependent: they grow exponentially. This means
 we are not strictly permitted to deploy the familiar
techniques of linear analysis. In particular, small perturbations cannot be,
in general, Fourier-decomposed in time. Modal solutions of the type
$e^{\sigma t}$ exist only
 when the growth rate $\sigma$ is much larger than
the growth rate $s$ of the channel solution itself, because then, as far as
the mode is concerned,
 the
background is steady (GX).
This is the case when the
amplitudes of the channels are large, as a consequence of the scaling $\sigma\sim
b\Omega$. So when
$b\gg 1$, it follows that $\sigma\gg \Omega\sim s$.

In order to make progress,
 we assume the limit of large channel amplitude and search for parasitic
 modes of type $\propto e^{\sigma t}$. The task then simplifies, requiring the
 solution of a
 boundary value problem in one variable with eigenvalue $\sigma$.

\subsubsection{Linearised equations}

The channel solutions outlined in the previous section are perturbed by a small
disturbance so that
$$ \v=2Ax\,\ey + \v^\text{ch}+\v', \quad   \B= B_0\,\ez + \B^\text{ch} +\B',\quad
P=P_0+P',$$
where the prime denotes the small perturbation. These are subtituted into Eqs
\eqref{u}, \eqref{B}, and \eqref{inc} which are linearised in the
perturbations. The next step is to assume that the channel amplitudes
are large, i.e. $b\gg 1$, and that the time dependence of the perturbation
scales as $\d_t \sim b\Omega$. This renders
 the background shear, Coriolis force,
uniform vertical field, and the exponential growth of the channel solution
all negligible. The channel structure is hence reduced to a stationary
equilibrium and so (within the assumptions made)
we are permitted to search for Fourier modes of the type
 $\v',\B',P' \propto e^{\sigma t + i k_x x + i k_y y}$. The analysis
reduces to an 8th order boundary value problem in $z$ with eigenvalue
$\sigma$. Its details are now sketched out.

Units are chosen so that $K=1$ and $b\Omega =1$. The velocity, magnetic,
and pressure perturbations are scaled
by $bv_0$, $bB_0$, and $bv_0^2$ respectively.  The linearised equations
 read as
\begin{align} \label{lineq1}
\sigma\, \v' &= -V\left\{\, v_z'\,\d_z
  \v^0+ i(\kk\cdot\v^0)\,\v' + (i\kk+\ez\,\d_z) \Psi'\,  \right\} \notag \\ 
& \hskip1.75cm + \frac{V}{M_A^2}\left\{ B_z'\,\d_z\B^0 +
  i(\kk\cdot\B^0)\,\B' \right\}, \\
\sigma\,\B'& = V \left\{\, B_z'\,\d_z\v^0 - u_z'\,\d_z \B^0 \right. \notag
  \\ &\left.
+ \, i(\kk\cdot\B^0)\,\v'- i(\kk\cdot\v^0)\,\B'\, \right\}
 - \frac{1}{b\Rm}(k^2 -\d_z^2)\B' \\
 0 &= i\kk\cdot\v' + \d_z v_z', \label{lineq2}
\end{align}
where the total pressure is $$\Psi'= P' + \frac{|B'|^2}{8\pi\rho_0},$$
 the scaled channel velocity is
 $V= v_0 K/\Omega$, which can be computed from Eq.~\eqref{v0GX}, the
planar wavenumber is defined by $\kk=(k_x,k_y,0)$, and $M_A$ is the Alfv\'{e}nic
Mach number.
 Finally, the background equilibrium fields are
given by
\begin{align*}
& \v^0= \sin z\,(\ex\,\cos\theta+\ey\,\sin\theta), \\
& \B^0= \cos z\,(\ex\,\cos\phi+ \ey\,\sin\phi).
\end{align*}

The linear operator associated with this problem is $2\pi$-periodic in $z$,
and we thus make the Floquet ansatz:  
$$\v'= \widetilde{\v}(z)\, e^{i k_z z}, \qquad \B'= \widetilde{\B}(z)\, e^{i k_z
  z}, \qquad
\Psi'= \widetilde{\Psi}(z)\, e^{i k_z z},$$
where the functions $\widetilde{\v}(z)$, $\widetilde{\B}(z)$,
$\widetilde{\Psi}(z)$ are $2\pi$-periodic in $z$. The Floquet exponent $k_z$ denotes
 the wavenumber of the perturbations' vertical `envelope', i.e.\ the large-scale
 structure greater than the channel width. The envelope wavenumber
  $k_z$
 takes values between
  $0$ and $0.5$. If $k_z$ extends beyond this range then we repeat, or
  `alias', eigensolutions which appear when $0<k_z<0.5$.
The Floquet
 decomposition reduces the problem
 to the domain $[0,2\pi]$, one channel wavelength, and furnishes us with the
 appropriate boundary conditions: the tilde eigenfunctions must
 be periodic. 

Now the problem is fully determined.
Given the parameters $A/\Omega$,
$\theta$, $\Rm$, and
$b$, on one hand, and $k_x$, $k_y$, and $k_z$ on the other we can compute
the eigenfunctions of the parasites and their growth rates $\sigma$.
Usually, we replace $k_x$ and $k_y$
 with the planar
wavenumber magnitude $k=\sqrt{k_x^2+k_y^2}$ and orientation angle
$\theta_k=\arctan (k_y/k_x)$. Also, for most occasions $\theta=\pi/4$ which
corresponds to the fastest growing channel in ideal MHD. Other orientations were tried but
we observed no change in the qualitative features of the results. A more
meaningful parameter, therefore, may be $(\theta-\theta_k)$ rather than $\theta_k$
itself. This angle difference measures the relative orientation of the mode's wavevector to the
fluid flow, and hence the degree to which it can extract shear energy, on one hand, and the
degree to which it escapes magnetic tension, on the other.

 In GX, the linearised equations could
be worked into a neat second order ODE by shifting to Lagrangian variables. Unfortunately, the 
resistive term prevents us from attempting a similar trick: the magnetic field
is no longer frozen into the fluid and as a consequence the induction equation
cannot be immediately integrated.
Instead, we solve the set of equations numerically via a pseudospectral technique. The
$z$-domain is partitioned into $N$ grid points and the operator
\eqref{lineq1}-\eqref{lineq2} discretised, with the derivatives computed using
Fourier cardinal functions (see Boyd 2000). A $7N\times 7N$ matrix
equation ensues taking the form of a generalised algebraic eigenvalue problem;
this 
 yields a spectrum that approximates
that of the original differential operator. The eigenvalues $\sigma$ are obtained using
standard linear algebra routines (the QZ algorithm, inverse iteration, etc). In the next few
subsections we present the results: in the ideal MHD case, and
with finite $\Rm$. But first, we discuss the interpretation of the
parasitic modes.

\subsubsection{Mode identification}

 In GX two classes of parasitic mode were shown to assail a
 developed MRI channel solution.
They were named Type 1 and Type 2 modes. The first was identified as the familiar
 Kelvin-Helmholtz instability, but the nature of the second was a little more
 difficult to assess. In fact, the Type 2 mode can be
 understood in the context of MHD jets
as something of a hybrid mode, consisting of motions associated with 
both the kink and pinch modes
 (Drazin and Reid 1981, Biskamp 2000). In this subsection we sketch out this idea.

In the fluids literature the classic jet profile takes the form
  $\text{sech}^2(z)$, the so-called `Bickley jet' (Bickley 1937), though other
  jet profiles offer the same qualitative behaviour (Drazin and Reid 1981).
 The Bickley jet is
  susceptible to two
classes of growing mode that have picked up a variety of names: there are the
 `kink' (or `odd', `Kelvin-Helmholtz', `sinuous') modes, and the `pinch' (or `even',
 `sausage', `varicose') modes. The names arise because the former
 buckles (or kinks) the jet vertically, while
the other squeezes and expands adjacent regions into a
pattern akin to a sausage. Quite generally, the kink mode grows faster than
  the pinch mode.
  But when a parallel magnetic field is present the kink mode can be
suppressed by magnetic tension which resists the mode's attempt to bend the jet and its embedded
 magnetic field lines (Chandrasekhar 1961). On the other hand, the pinch mode can tap
into the free energy locked up in the magnetic configuration, to some extent,
 and enhance its growth (Biskamp et al.~1998, Biskamp 2000). 
If field lines are permitted to break and reconnect (i.e. if $\Rm$ is finite)
 further energy
is available to the pinch mode and it adopts some of the characteristics
 of a tearing mode.

The Type 1 and Type 2 parasitic modes of the MRI channel problem are
the generalisations of the kink and pinch modes to a situation comprising 
an infinite periodic lattice of countermoving jets. As such, they are global 
 modes, attacking \emph{all} the channels concurrently and are hence able to exhibit
 structure larger than one channel width.  This larger scale structure is
 summarised by the Floquet factor $e^{i k_z z}$. In comparison, the classical pinch and kink
 modes are local modes which attack an \emph{isolated} channel. (In real
 accretion disks, this `global structure' --- and the structure of the channel
 itself --- will be influenced by the
disk's vertical stratification and boundary conditions.) 
The Type 1 mode we interpret as the straightforward
  generalisation of the kink mode; it is a stationary instability and buckles
  both sets of inward and outward jets concurrently. The Type 2 mode is a
  little more complicated.
 We interpret it as a hybrid kink and pinch
  mode: each Type 2 mode `kinks' one set of jets (either the inward or the
  outward sets) while `pinching' the other
  set of jets. There are hence two Type 2 modes and their growth rates are complex
  conjugates, meaning they manifest as travelling waves. 
  We sometimes refer to the Type 2 mode as `the kink-pinch mode'. 

The key parameter which distinguishes between the two types of parasite is the
Floquet exponent, $k_z$. The generalised kink mode is characterised by small $k_z$
and the pinch mode by larger $k_z$. The transition between the two, as $k_z$
is varied, is rapid but smooth.

\subsubsection{Ideal MHD}

We begin by reexamining the ideal MHD case, setting the resistive terms to
zero.
The problem now returns to that tackled by GX; we briefly restate some of their
results. 

First, we exhibit the analytic stability criteria
derived by GX which are derived
in the long horizontal wavelength limit. Because these long modes are the
most susceptible to instability they give conditions on stability generally.
When $k_z=0$, which corresponds to a classic Type 1 instability, or `kink'
mode, stability is assured if
\begin{equation} \label{kstab}
|\theta-\theta_k|> \arctan M_A.
\end{equation}
Note that the Alfv\'{e}nic Mach number is a complicated function of $\theta$ but is always
$\lesssim 1$. Therefore, a very rough, but revealing, estimate of stability
is $|\theta-\theta_k|>\pi/4$. For the $k_z=0$ kink mode to grow, its wavevector must
be sufficiently aligned with the jet velocity, within some angle less than 45
degrees. The more the wavevector $\kk$ points away from $\v^\text{ch}$, the less
shear energy is available and the greater the magnetic tension. Two-stream
channels appear regularly in numerical studies and in such simulations
only $k_z=0$ kink parasites can fit into the computational domain. Thus the
criterion \eqref{kstab} is the key one to keep in mind.

 In contrast, when $k_z$ is non zero (a parameter choice
 associated with some Type 1 kink modes and
 all the
Type 2 modes) the stability criterion is more stringent: these
 modes grow more easily. We have stability when
\begin{equation} \label{pstab}
 |\theta-\theta_k| > \arctan (M_A/3),
\end{equation}
from which the rough estimate $|\theta-\theta_k| > \pi/3$ can be derived as above.
Such modes grow for a wider range of orientations. Somehow the global
structure, absent in the $k_z=0$ case, allows them to better resist magnetic tension. 

The two types of mode discovered by GX can be distinguished by the parameter
$k_z$. If $k$ and $\theta_k$ are fixed and we vary $k_z$ from $0$ to $0.5$, the
mode transforms smoothly (though not uniformly) from the Type 1 kink mode to
the Type 2 kink-pinch mode.  
When  $0 \leq k_z \lesssim 0.2$ we obtain kink modes, and
 when $0.2 \lesssim k_z  \leq 0.5$  we
 obtain kink-pinch modes. The actual values, however, depend somewhat on $k$
 and in
 a complicated fashion, so we omit these details. We also omit the graphs of
 the growth rates as functions of $k$, as these appear in GX. In summary, Type 1
 modes grow at a rate an order of magnitude larger than Type 2 modes with the
 maximum rate at values near $k=1/2$. Neither class of mode can grow when
 $k \geq 1$.

In the following four figures we plot representative examples of the two classes
of mode. We set $\kk$ and $\v^\text{ch}$ parallel to each other,
by enforcing $\theta_k=\theta=\pi/4$.
At this orientation both kink and pinch modes enjoy
 the greatest access to the shear energy of the
background flow. At the same time, they are completely free of magnetic tension,
being perpendicular to $\B^\text{ch}$. Consequently, these modes are virtually
hydrodynamic, consigning
their magnetic perturbations to the total pressure.

 Fig.~1 shows three components of the kink mode eigenfunction, for
which $k_z=0$. Note that the total pressure perturbation $\widetilde{\Psi}$ changes sign
at the two altitudes $z=\pi/2 $ and $3\pi/2$, which are where the two magnetic null
surfaces are located, and to where the fluid jets are
concentrated. Hence a pressure gradient exists across each jet and
this gives rise to a buckling,
or kink, motion. This motion can be observed in the $\widetilde{u}_z$
perturbation, which possesses maxima and
minima at the jet centres: the jets are deflected upward or downward. To
emphasise this point, we present in Fig.~2 the real part of the $x$ component of $\v^\text{ch}+\v'$
 in the $(x,z)$ plane.
The background is a four-stream channel and we show two periods (in $x$) of
the velocity. The kinking of the four
 inward and the four outward jets is pronounced, as is the formation of
nascent Kelvin-Helmholtz type billows.

Fig.~3 shows three components of the Type 2 mode for the same parameters save
$k_z$, which takes the value $0.5$. If we turn our attention first to the
altitude $z=3\pi/2$ we notice that the eigenfunction behaves similarly to
the kink mode: the total pressure perturbation changes sign at the
centre of the jet and $\widetilde{u}_z$ possesses a maximum there. However, the companion 
jet at altitude $z=\pi/2$ undergoes a different kind of perturbation. Here it
is the $\widetilde{u}_z$ perturbation which changes sign; accordingly, fluid above and
below the jet will be either attracted to the centre of the jet or
repelled. This is characteristic of the pinch, or sausage, motion.
 In Fig.~4 the $x$-component of the perturbed four-stream channel velocity is
presented, as before, in which we can observe the alternate kinking and
pinching of neighbouring jets. There exists a companion Type 2 mode
 with complex conjugate growth rate, and its eigenfunction reverses the kink and pinch motions to
$z=\pi/2$ and $3\pi/2$ respectively.

\begin{figure}
\scalebox{.55}{\includegraphics{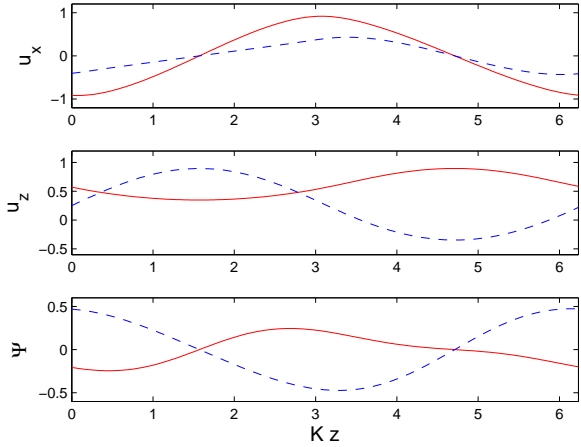}}
\caption{The $\widetilde{u}_x$, $\widetilde{u}_z$, and
    $\widetilde{\Psi}$
 components of the eigenfunction
    of the kink mode with
  $\theta=\theta_k=\pi/4$ and $k/K=0.5$ and $k_z=0$. The solid line represents the real
 part and the dashed line represents the imaginary part. Note that the pressure
  pertubation $\widetilde{\Psi}$ changes sign at $Kz=\pi/2$ and $3\pi/2$ which are the
  altitudes at which the two jets are concentrated. A kinking or buckling of
  the jet is the outcome. Because $\theta=\theta_k$ the mode is effectively
  hydrodynamic. The total eigenfunction is normalised so that
 $\text{max}|\tilde{u}_x|=1$. The growth rate is $\sigma/(b\Omega)=0.1959$.}
\end{figure}
\begin{figure}
\begin{center}
\scalebox{.55}{\includegraphics{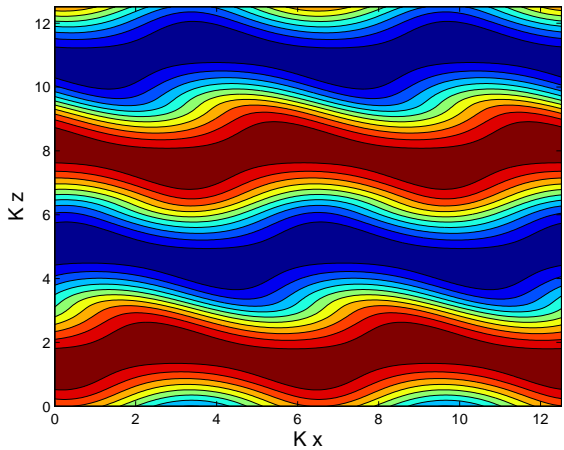}}
\caption{Coloured iso-contours of the real part of
    $v^\text{ch}_x+v_x'$ at $y=0$ for the kink mode in the $(x,z)$ plane.
  Parameters chosen are
 $\theta=\theta_k=\pi/4$, $k_x/K=0.5$, and $k_z=0$. The background is a
    four-stream MRI channel with jets centred at $Kz=n\pi/2$ with $n=1$, $3$,
 $5$, and $7$. The perturbation is normalised so that $\text{max}|u_x'|$ is a
  a quarter of $\v^\text{ch}$. }
\end{center}
\end{figure}

\begin{figure}
\begin{center}
\scalebox{.55}{\includegraphics{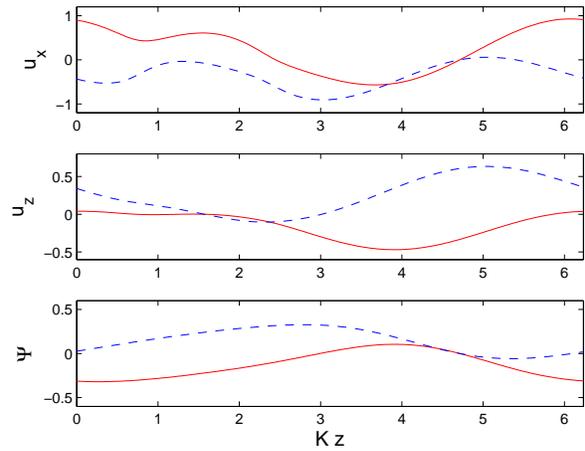}}
\caption{The $\widetilde{u}_x$, $\widetilde{u}_z$, and
    $\widetilde{\Psi}$
  components of the eigenfunction
    of the kink-pinch mode with $\theta=\theta_k=\pi/4$ and $k/K=0.5$ and
    $k_z/K=0.5$. Note that the pressure perturbation $\widetilde{\Psi}$ changes sign at $Kz=3\pi/2$,
    giving rise to a kinking motion upon the jet centred there,
 while the vertical velocity perturbation
    $\widetilde{u}_z$ changes sign at $Kz=\pi/2$ giving rise to a pinching motion upon the
    jet centred there. Because $\theta=\theta_k$ the mode is effectively
  hydrodynamic. The growth rate is $\sigma/(b\Omega)=0.0931-0.1319\,i$. }
\end{center}
\end{figure}
\begin{figure}
\begin{center}
\scalebox{.55}{\includegraphics{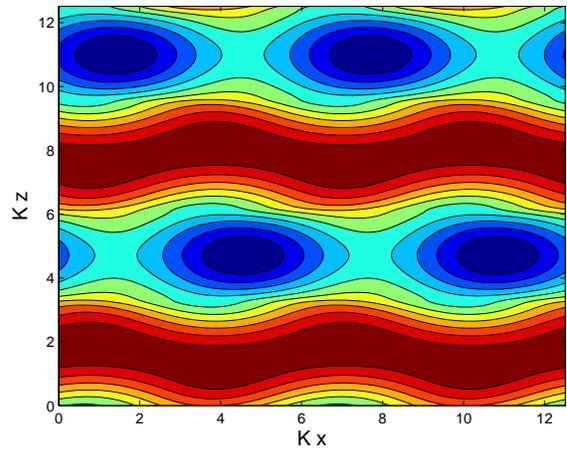}}
\caption{Coloured iso-contours of the real part of
    $v^\text{ch}_x+v_x'$ at $y=0$ for the kink-pinch mode in the $(x,z)$ plane.
  Parameters chosen are
 $\theta=\theta_k=\pi/4$, $k_x/K=0.5$, and $k_z/K=0.5$. The background is a
    four-stream MRI channel with jets centred as before. Note the alternate
  kinking and pinching of the jets. Also the entire pattern is moving to the
  right because $\sigma$ possesses a negative imaginary component. }
\end{center}
\end{figure}

Next we examine the eigenmodes for different wavevector orientations. In
particular, we examine axisymmetric modes, $\theta_k=0$, when $\theta=\pi/4$.
At this configuration only modes with nonzero $k_z$ grow (Eq.~\eqref{kstab}),
and we limit ourselves to $k_z=0.5$.
In these cases the background magnetic field plays a larger role in the
dynamics, and leads to smaller growth rates and more
stable short scales. In Figs 5 and 6 we present six components of the eigenfunction
when $k=0.25$. In Fig.~5 the $\widetilde{u}_z$ and $\widetilde{\Psi}$ perturbations
are similar to their `hydrodynamical' $\theta_k=\theta$ cousins in
Fig.~3. Note however the extra structure exhibited by the $\widetilde{u}_x$ component near
$z=\pi/2$, $\pi$ and $2\pi$. These abrupt variations react to the contortions
of the $\widetilde{B}_x$
and $\widetilde{B}_y$ magnetic perturbations frozen into the fluid. But at
the pinch altitude $z=\pi/2$, the magnetic perturbations are precisely zero;
this means that the pinch motion here preserves the magnetic null surface of the
background channel flow. Here the channel's magnetic field lines are symmetrically
squeezed towards
 the null
surface, or symmetrically repelled. In contrast, at the kink altitude $z=3\pi/2$, the
null surface is distorted upward or downward alongside the field
lines.

\begin{figure}
\begin{center}
\scalebox{.55}{\includegraphics{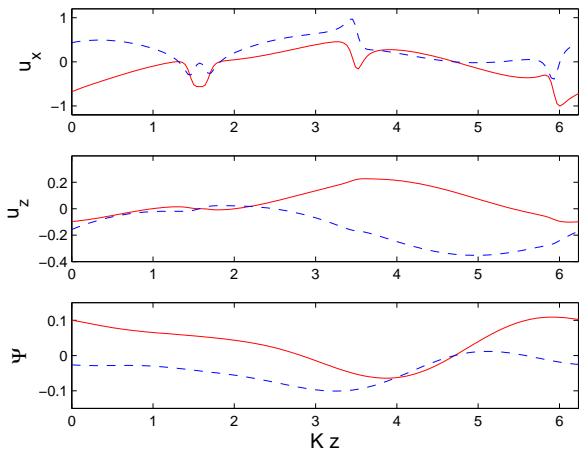}}
\caption{The $\widetilde{u}_x$, $\widetilde{u}_z$, and $\widetilde{\Psi}$ components of the
    eigenfunction of the kink-pinch mode when $\theta=\pi/4$ and $\theta_k=0$
    with $k/K=0.25$ and $k_z/K=0.5$. Resistivity is set to zero. 
The mode is hence axisymmetric and magnetic stresses play
    an important part in the dynamics. The pinch motion is located at
    $Kz=\pi/2$ and the kink at $Kz=3\pi/2$. The growth rate is $\sigma/(b\Omega)=0.0142-0.1216\,i$.}
\end{center}
\end{figure}
\begin{figure}
\begin{center}
\scalebox{.55}{\includegraphics{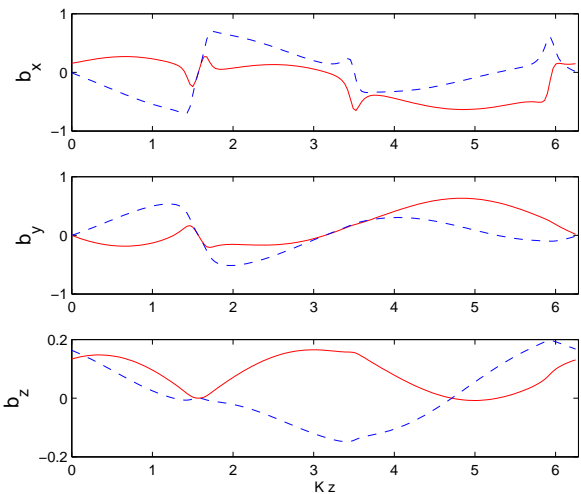}}
\caption{The magnetic components of the
    eigenfunction of the kink-pinch mode of Fig.~5 when $\theta=\pi/4$ and $\theta_k=0$
    with $k/K=0.25$ and $k_z/K=0.5$. Note that at the pinch at $Kz=\pi/2$ the components are all
 zero and thus the magnetic null surface of the channel solution is preserved.}
\end{center}
\end{figure}

 GX suspected that the pinch motion associated with the Type 2 mode would lead to reconnection at the
null surface when resistivity is present,
 and to the consequent formation of magnetic islands, or plasmoids.
 Such behaviour has been
seen in isolated jets by Wang et al.~(1988) and Biskamp et al.~(1998). 
We investigate this phenomenon
in the next subsection in the case of a periodic lattice of MRI channels.

\subsubsection{Resistive MHD}

The magnetic Reynolds number $\Rm$ is now set to finite values and the channel amplitude
$b$ enters as another parameter. A number of changes take place: (a) the
structure and growth rate
of the kink-pinch mode alters, and (b) new growing `pure-pinch', or
`pinch-tearing',
 modes emerge.
 Meanwhile,
 the kink mode's structure and growth rate do not
change significantly, nor those modes
that possess wavevectors perpendicular to $\B^\text{ch}$.

\begin{figure}
\begin{center}
\scalebox{.55}{\includegraphics{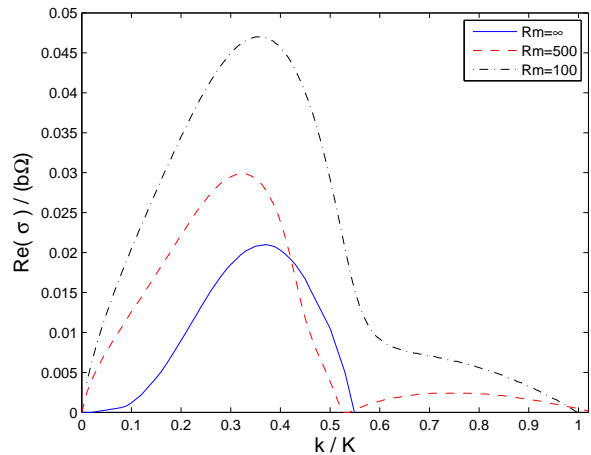}}
\caption{The real part of the growth rate $\sigma$ 
as a function of $k$ for an axisymmetric Type 2 mode ($\theta_k=0$)
 when $\theta=\pi/4$ and $k_z/K=0.5$ in the resistive and ideal cases. The solid
 curve corresponds to the ideal $\Rm=\infty$ case, the dashed line to the
 $\Rm=500$ and $b=5$ case, and the dotted-dashed line to the $\Rm=100$ and
 $b=5$ case. Note the larger growth rates and the extended range
 of instability when resistivity is present. Note also that for
 intermediate $\Rm$ there is a range of $k$ for which resistivity plays a
 `diffusive' damping role. Also be aware that changing $\Rm$ also alters 
  the underlying equilibria slightly in each case.}
\end{center}
\end{figure}

The resistive Type 2 modes achieve growth rates larger than their ideal
MHD counterparts, an increase which seems to scale like $(b\Rm/M_A)^{-3/5}$ (in
agreement with studies of the tearing mode, cf.\ White 1986).
 By removing the constraint of flux conservation, the mode can reconnect
magnetic field where it pinches the field lines. It thus gains
 access to extra energy stored in
the channel's magnetic configuration. Such energy is additional to that
extracted from the velocity shear and allows the mode to grow more
rapidly. These increases can be observed in Fig.~7 for an axisymmetric pinch mode
with $b=5$ and $\Rm=\infty$, $500$, and $100$. Resistivity also transforms the long
wavelength
 behaviour of the mode: its growth rate no longer scales as $k^3$ for small $k$; instead,
 $\sigma$ possesses a
 much steeper dependence on $k$ (some fractional power).
Conversely, instability emerges on shorter scales previously forbidden by
ideal MHD: the `second hump' in Fig.~7. Similar short-scale
behaviour has been reported by Biskamp et al.~(1998) for an
isolated two-dimensional jet but with $\B^\text{ch}$ parallel to
$\v^\text{ch}$. 

The $\Rm=500$ curve reflects more complicated behaviour at intermediate
$\Rm$ and deserves a few extra comments. On a small band
of $k$ near $0.5$ the ideal MHD mode grows \emph{faster} than the resistive
mode; in addition the large and small $k$ branches of $\sigma$ detach.
 Note that resistivity alters the
 background equilibria, so the two modes are not exactly analogous. But even if this
is taken into account, the resistive growth rates
 remain slightly lower for these values
of $k$. We conclude that resistivity's `diffusive role' is anomalously important on
this narrow range of intermediate scales and counteracts the extra growth
 stimulated by reconnection. When resistivity is
sufficiently strong ($\Rm\lesssim 250$) the effect vanishes.

The eigenfunction of the resistive kink-pinch mode does not differ significantly
from its ideal MHD counterpart. The most notable distinction is in the
magnetic perturbations which no longer preserve the magnetic null surface at
the pinch altitude (cf.\ Fig.~6). Instead, a chain of magnetic islands are
formed. (In the interest of brevity these profiles are not presented.)

\vskip0.3cm

 Resistivity stimulates the growth of new `pure-pinch' or `pinch-tearing' modes.
 These extract little to no
 energy from the velocity shear their growth relying solely on the reconnection
 of magnetic field. Consequently, they favour wavevector
 orientations along or near $\B^\text{ch}$ and do not grow at all when $\kk$ is
  perpendicular to the magnetic field. Typically the growth rates are small
  with $\sigma\sim 0.01$, though they can appear for \emph{all} $k_z$ (unlike the
 kink and
 kink-pinch modes).

\begin{figure}
\begin{center}
\scalebox{.55}{\includegraphics{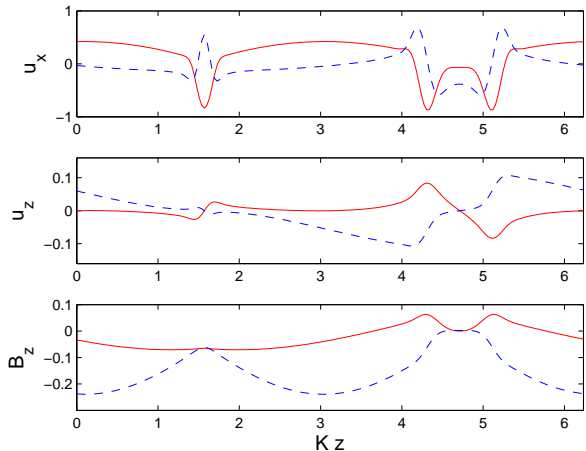}}
\caption{Selected eigenfunction profile
  for a general pinch-tearing
    mode.
    Parameters are $\theta=\pi/4$, $\theta_k=-\pi/6$, $k_z=0$
    and $k/K=0.4$, while $\Rm=500$ and $b=5$. The growth rate is 
$\sigma/(b\Omega)= 0.0096$. Though the mode
    pinches both jets concurrently it localises preferentially on the upper
    one. A companion mode, growing at the same rate, localises on the lower jet.}
\end{center}
\end{figure}

Their fastest growth is achieved, naturally, when $\theta_k=\phi$. At this
orientation the structure of the mode is akin to the `double tearing mode' --- a
pinching motion upon each channel (Pritchet, Lee, and Drake 1980).  That being said,
the growth rates of modes oriented this way
scale at the slower rate of the single tearing mode: $\sigma\sim (b\Rm/M_A)^{-3/5}$. 
When $\theta_k$ departs from $\phi$ the mode structure 
and growth rate change rapidly. The pinch-tearing mode splits into two modes,
each localised primarily to one or the other channel. In addition, the growth rates
decrease and pick up imaginary parts. In Fig.~8 we plot a representative
eigenfunction. At larger $k_z$ these modes
generally do not appear for all $\theta_k$;
 they are more likely to be supplanted by the
resistive Type 2 modes.
 Pinch-tearing modes, like the kink and kink-pinch modes, grow
only on long lengthscales. When $k \geq 1$ they decay. 

\vskip0.3cm

As the reader may have noticed, the kink modes, kink-pinch modes, and the
pinch-tearing modes paint a rather complicated picture as
we vary the governing parameters, especially for intermediate $k_z$.
 The main points
 to take away, however, may be summarised neatly. Resistivity aids in the
 destabilisation of a channel solution by exacerbating ideal instabilities
 (the Type 2 mode), and by introducing the pinch-tearing instability.
 Resistivity allows unstable modes for \emph{all} combinations
 of $\theta_k$ and $k_z$; unlike
  ideal MHD, growth need not be restricted to orientations in a sector
 around $\theta$. The only restriction that remains is on the magnitude of $k$
 (which must be less than $1$).
 Still, the fastest growing mode remains the
 hydrodynamical kink instability. The other slower
 instabilities may be important in simulations where the geometry of the
 computational domain excludes certain wavevector orientations and hence
 certain growing modes.

\subsection{Discussion}

 Our main assumptions so far have been
  incompressibility, and that
 $b \gg 1$ (i.e.\ the condition that channels are well
developed).
Unfortunately, these two assumptions may not be consistent in practice.
 Once the channel reaches the regime $b \gg 1$, 
and our \emph{modal} incompressible analysis applicable, the magnetic
 pressure and thermal pressure become comparable. But this
  may render compressibility important, and our \emph{incompressible} modal analysis
 inapplicable~! We hasten to add that
  the preceding work may still make adequate
 qualitative predictions in the $b\gg 1$ (and perhaps the $b\sim 1$) regime, even if 
 the quantitative details go astray. 
Compressible shear modes should emerge and attack narrow compressible channels 
in ways analogous to their incompressible counterparts. On the other hand, at
low channel amplitudes `Kelvin-Helmholtz processes', be they kinking or
pinching, should be present, even if they can no longer take modal form 
(i.e. be $\propto e^{\sigma t}$).

 We examine these two regimes in more
 detail, the $b\sim 1$ regime with qualitative arguments, and the $b\gg 1$
 regime by solving another boundary value problem (the full details of which
 we put in an appendix).

\subsubsection{The incompressible $b\sim 1$ regime}

When $b\sim 1$ the situation is
complicated by the rotation and shear issuing from orbital motion, on one
hand,
 and the exponential growth of the channel solution itself, on the other. If we were
to attack the problem directly we would reframe the linear perturbation 
equations in shearing
coordinates (Goldreich and Lynden-Bell 1965) and then massage them into an
initial value problem consisting of 6 coupled PDEs in $t$ and $z$. To then compute
a numerical solution is a
somewhat laborious task, which we decline, but a few qualitative points can be
made. 

The shearing background's main effect is in introducing a
time-dependence to the pertubation's wavevector. Specifically, the modes'
orientation angle $\theta_k=\theta_k(t)$ should decrease from positive to
negative values with $t$ (corresponding to clockwise rotation of $\kk$). First
consider
how this would impact on the ideal MHD modes. If we apply,
naively, the criterion that perturbations grow only when their $\theta_k$ is
within some range encompassing $\theta$ (the orientation of the channel) then
we would expect a burst of growth as $\theta_k(t)$ passes through a
neighbourhood of $\theta$, and little change before and afterwards. The
amplification of the perturbation during this period may be sufficient to
disrupt the channel.
 This bursting behaviour is observed in GX's
solution to the inital value problem. They find a short period of exponential growth
in a region encompassing $\theta_k=\pi/2$, which makes sense because the
background they chose was a marginal
MRI channel with $\theta\approx \pi/2$. 

When resistivity is added, however, there should
be growing perturbation for \emph{all} values of $\theta_k(t)$, because of 
 the new pinch-tearing
process which favours those orientations ignored by the kink (and kink-pinch)
mode. 
Earlier we showed that as pinch-tearing modes, this process
grows at a rate of order  $0.01\, b\Omega$, significantly less than the kink parasites.
But if we simply extrapolate to the $b\sim 1$ regime, then 
the analogous modes will grow significantly less than the rate of change of
the wavevector itself. 
In the time the
 wavevector $\kk$ sweeps through the sector of favourable growth, the
 pinch-tearing instability will probably
 not have grown sufficiently fast to disturb the channel.
 We, as a consequence, envisage more or less the same
 behaviour in the ideal and resistive cases.

Much more important than the shearing background
 is the role of the channel's exponential increase. 
We argue that the rapid growth of the channel necessarily \emph{delays} the advent of the
 parasitic instabilities until the channel achieves large amplitudes.
A small perturbation on a low amplitude channel ($b \sim 1$)
grows at a rate $\lesssim\Omega$, a similar rate to the channel itself. But
 this means that
 as time progresses the amplitude of the parasite relative to the
channel \emph{will remain small}: an MRI channel can `outrun' its
(small amplitude)
parasites, simply by virtue of growing at a similar rate. When
it reaches the regime $b\gg 1$ only then can the parasitic growth overtake
the channel and overwhelm it.  It follows that the
 regime $b\gg 1$, and the strong magnetic fields associated with it, 
 are
 natural outcomes of the channel formation mechanism. As we
 show a little later, some simulations reveal this behaviour,
 and aside from numerical applications, this may be of relevance to observed singular flaring
 events.

\subsubsection{The compressible $b\gg 1$ regime}

The delayed onset of the parasitic modes
 allows the MRI channels to achieve large amplitudes before they collapse. 
But, as mentioned in Section 2.1,
the strong magnetic pressure associated with $b\gg 1$ emperils the
  incompressibility assumption. In addition,
channel structure will depart significantly from the simple sinusoidal profiles of
 \eqref{ch1} and \eqref{ch2} and its time dependence will not be limited to
 the amplitude factor $e^{st}$. For these reasons the analysis of
 Section 2 offers only the most general guide. 

To understand channel breakdown in this regime more fully,
we conducted a linear stability analysis of `extreme'
compressible channels. For simplicity 
we assumed that the disk was composed of an 
isothermal ideal gas, characterised by the sound speed $c_s$. 
For the channel profiles, we constructed convenient
 analytic approximations motivated by profiles observed in 1D
 simulations (unfortunately, there is not a straightforward
 way to calculate these profiles \emph{a priori}).
See Appendix A
for the detailed analysis. Here we offer a brief summary.

\begin{figure}
\begin{center}
\scalebox{.5}{\includegraphics{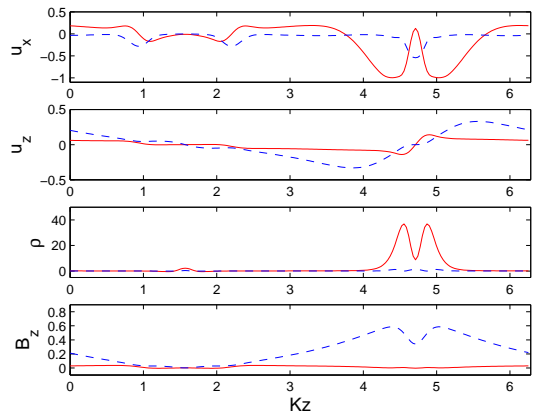}}
\caption{Selected eigenfunction components of the compressible
pinch-tearing mode localised on the upper jet. Parameters of the equilibrium are  $M_A=0.7$,
    $\beta=0.1$, $d=5$ (see Appendix A), and the perturbation is determined by
  $\theta_k=-\pi/4$ and $k_x/K=1$. The growth rate is 
$\sigma/(v_0K)= 0.1041 +0.6982\,i$. An equivalent mode exists, growing at the same rate,
    but is localised to the lower jet.}
\end{center}
\end{figure}
\begin{figure}
\begin{center}
\scalebox{.5}{\includegraphics{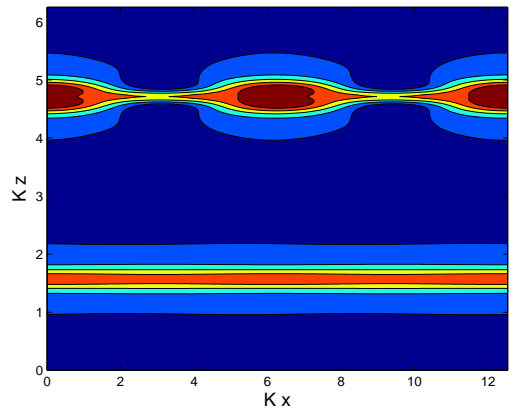}}
\caption{Coloured isocontours of the total density of the
    compressible pinch-tearing mode in Fig.~9 in the $(x,z)$ plane. The perturbation is normalised
    so that it is a fifth of the equilibrium density profile. Two periods in
    the $x$ direction are included.}
\end{center}
\end{figure}

The most important instabilities we found
 were the kink mode and the pinch-tearing mode. Importantly, their relative 
behaviour changes in interesting ways
 as the  plasma beta parameter $\beta\equiv 2c_s^2/ v_A^2$ varies from
 large values (the incompressible regime) to low values (narrow compressible channels).
In summary, as $\beta$ shrinks (a)
 instability occurs on significantly shorter horizontal lengthscales, (b)
the unstable modes generally localise on one or the other jet, (c) the pinch
 mode growth rate increases significantly and can reach levels comparable to 
the kink mode. For certain parameter choices the pinch mode can be the
 dominant mechanism of channel destruction. In Figs 9 and 10 we plot a
 representative example of a pinch mode on an extreme compressible channel, so
 as to compare with our numerical simulations in Section 3.

\section{Compressible simulations}

So far we have attacked the problem of the MRI's nonlinear behaviour
by reducing it to two linear calculations: that of the MRI channel
flow, and that of the parasitic modes which feed on it. 
In essence, we have concentrated on the life of a single MRI mode and
its associated nonlinear solution, while
ignoring its possible
nonlinear interactions with other MRI modes. 
 In reality, such interactions are inevitable and may prevent the
 formation of the pure nonlinear channel structures upon which we have
 based our analyses. In the previous section, we argued that `pure' isolated channels
 will grow naturally to enormous amplitudes before they collapse --- the fact that simulations do
 not always report this suggests that nonlinear effects early in
 channel growth are important.

In this section we numerically simulate the equations of nonideal MHD in order
to isolate the
emergence of parasitic instabilities in various contexts. We then
 compare the observed behaviour to the predictions of the preceding
linear analyses.
 Concurrently, we discuss the role of the parasites more broadly and make a few comments
about the saturation of the MRI and the role of the computational
geometry. But first, we review the
simulation literature and some conclusions we can draw from it.

\subsection{Brief review}

Channel flows were discovered in the very first axisymmetric simulations of
the MRI in a shearing box (Hawley and Balbus 1992). The box would
 be dominated initially by the fastest growing MRI channel,
 but if the vertical wavelength of this solution was smaller than the
vertical dimensions of the box the system would witness an inverse cascade to
longer wavelength channel flows. (The catalyzing process is perhaps the
axisymmetric Type 2 parasitic instability --- but see also Tatsuno and Dorland
2008.) Later
 fully three-dimensional simulations revealed that analogous channel solutions
 are a recurrent feature of MRI saturation --- but only in the case when the
 imposed $\B$ field exhibits a net flux. Zero net flux runs have never 
exhibited these coherent structures. Notable works which simulate shearing boxes
with a uniform $B_z$ are
Hawley et al.~(1995) and Sano et al.~(2004), who work with the
equations of compressible ideal MHD, Fleming et al.~(2000), Sano and Inutsuka
(2001), and Sano (2007), who add resistive dissipation explicitly, and Lesur
and Longaretti (2007), who undertake incompressible runs with both 
resistivity and viscosity.

 All these studies employ the classic `bar'
computational domain, a box with dimensions $L_x\times L_y\times L_z$,
typically elongated in the $y$-direction so that $L_x=L_z=L$ and $L_y= 4L$ or
$2\pi L$. In this geometry simulations exhibit a pattern of irregular
and recurrent
`bursting' events whereby two-stream channels form and then collapse in
turbulence. For the most part, the amplitudes of these structures
 are never particularly large, possessing a magnetic pressure of order
 the thermal pressure at the peak of channel formation. In between the
 peaks, the disordered troughs possess a magnetic pressure of some tenth
  the thermal pressure.
 The troughs are never so low that the linear MRI modes
 decouple, which means the problem remains stubbornly nonlinear. Exceptions
 occur when the system is near criticality or, sometimes, in the initial stages of the
 evolution; in these cases clean isolated channels emerge and reach large amplitudes
 (Lesur and Longaretti 2007, Section 3.2).

Sano et al.~(2004) show that the larger the initial $\beta$ the less
prominent the intermittent channel formation; both the amplitudes and
frequency of channels drop. In fact, when the imposed field is sufficiently weak ($\beta$
very large) the channels fail to appear at all. The same trend has been shown
by Bodo et al.~(2008) when they increase the radial size of the computational
domain. In both cases (larger $\beta$, larger box) more active MRI modes can
fit into the computational domain, and hence a richer nonlinear dynamics is
available. We suspect, as a consequence, that the nature of the recurrent
channel behaviour is governed by nonlinear interactions between MRI modes, as
opposed to linear parasitic behaviour, even if it is usually claimed
 that parasitic modes (as described in GX and Section 2) are responsible for the channels'
 demise. This is explored in more detail in the following subsection.

\subsubsection{Channel breakup in bar geometry}

 We restrict attention here to simulations of recurrent low amplitude channels
 in the `bar' geometry.
The first thing to note is
that, according to the incompressible analysis of Section 2, most
two-stream channels are \emph{stable} to the kink mode.
Generally, the two-stream channel possesses orientation $\theta \lesssim
\pi/4$, and kink modes that are both properly oriented, cf.~\eqref{kstab},
 and which can
fit into the computational box, are of
 too
short a wavelength to be unstable (they must have $k<K$).
 The only exceptions
are two-stream MRI channels near marginal stability 
 (as in Fleming et al.~2000 and Lesur and Longaretti 2007) for which
 $\theta\approx \pi/2$.
On the other hand, two stream channels are unstable to the 
 $k_z=0$
pinch-tearing mode.
It is likely though, owing to their small growth rates ($\sim 0.01\,b\Omega$),
 that these modes play little role in the dynamics when $b\sim
 1$. 
They should be swamped
 by the growth of the channel itself, which is at least an order
 larger. Moreover, simulations of
 recurrent channels show the breakup takes place on the
orbital timescale, which is much shorter,
 and does not proceed by the formation of plasmoids, etc. Lastly, 
 modes with nonzero
 $k_z$ cannot fit into the computational domain.

 One, of course, could argue that the simulated channels possess $b\sim 1$
 and hence are
 outside the jurisdiction of Section 2's analysis (which requires $b\gg
 1$).
It may be that shorter wavelength kink instabilities ($k>K$) exist
 in this regime.
But even if this is allowed for, like the resistive pinch modes, nascent
 kink modes cannot grow faster than their hosts and would therefore 
not provide a viable
 way to disrupt the channel. Indeed, we have simulations in larger domains
 of clean growing
 channels of low amplitude that remain undisrupted even when vulnerable to
 kink modes (see Section 3.2.1)\footnote[1]
{When $b\sim 1$ another problem arises from fundamental difficulties in
 relating predictions based on the
  shearing sheet (which possesses an infinite
 domain) to simulations performed in the shearing box
 (which enforces shearing periodic boundary conditions). For instance, in the
 former, Eulerian $k_x$ can achieve arbitrarily small values while, in the latter, $k_x$
 is limited by the size of the radial domain. In fact, x-periodic Eulerian
 modes are not well-defined in shearing boxes, except on timescales short
 compared to the shear time. These difficulties are not crucial 
to our principal points.}

So how are low-amplitude recurrent channels destroyed? 
 The critical issue, in our opinion, is
  the fact that the repeated two-stream channels emerge from a disordered turbulent
trough which is \emph{outside the linear regime}. For this reason the `background' harbours
other MRI modes of appreciable amplitudes which
 are competing against, and interacting with, the two-stream
channel. The two-stream flow eventually dominates but it is
highly `contaminated' by its MRI counterparts: perturbations upon it have already 
achieved nonlinear amplitudes and the clean linear analysis of Section 2, nor
a generalisation of it to the $b\sim 1$ regime, is valid.
 Perhaps it might be fruitful to instead consider the nonlinear stability of
 the
 two-channel structure in this situation: short wavelength kink disturbances in this
 regime may be unstable, in contrast with the linear analysis.
 Alternatively, we could regard the channel's destruction simply as a result of
 turbulent mixing. The eddy turnover time of the MRI-induced turbulence, after
 all, is of order $\Omega^{-1}$ and hence similar to the growth time of the channel.
 The latter interpretation is supported by the system's response to
   changes in $\beta$ and box size.
  Increasing either introduces more unstable MRI modes into the box, which enrich the nonlinear
 turbulent dynamics in the troughs and hence impede the formation and
aid the destruction of two-stream channels. Admittedly, the precise details
 of how these new degrees of freedom
 work out in practice is not at all straightforward. 
But whatever the details are, we think it is incorrect to
simply attribute channel destruction to a linear `parasitic mode' along the lines of GX or
 Section 2.

\subsection{Numerical results}

We now put aside the interesting questions of saturation and concentrate a
little on the perhaps more mundane but also more precise and tractable task of exhibiting
and examining the parasitic modes as they appear in the simulations. The
preceding studies never clearly isolated and revealed the structure of the
parasites to which they often appealed. Once this is done, we briefly show
simulations of recurrent channels.

Our computations employed
a modification of \textsc{Zeus3D} (see Stone and Norman 1992a, b) to solve the
equations 
of nonideal, isothermal, compressible 
MHD in a shearing box, while assuming
an ideal gas law. Dissipation is modelled with an Ohmic resistivity and a
Navier-Strokes viscous stress. The equations and numerical set
up are described in Lesaffre et al.~(submitted). 

 Dimensions are chosen so that
$\Omega=1$, $\rho_0=1$, and $L_z=1$, where $\rho_0$ is the initial mass
density and $L_z$ is the vertical length of the box. The sound speed is $c_s$
 and is set to $\sqrt{5/3}$, except for one subsection where it is 
$10\sqrt{5/3}$.
The pressure
scale height is $H= c_s/\Omega$, and is thus roughly the vertical size of the
box in the first case, and more than ten times it in the second.
The other parameters are the initial Alfven velocity $v_A$, the
viscosity $\nu$, and the resistivity $\eta$. These take the values:
$\nu=\eta=2\times 10^{-3}$ and $v_A=\sqrt{2}/10\approx 0.141$. The plasma
$\beta$ is consequently $1.67\times 10^2$ (when $c_s=\sqrt{5/3}$), or
$1.67\times 10^4$ (when $c_s=10\sqrt{5/3}$).
Our choice of $v_A$ ensures that the fastest growing channel mode has wavenumber close to 
$K=2\pi/L_z$ and thus exhibits two-stream structure
 whatever $c_s$. No other growing channel
solutions fit in the box, though other $k_x\neq 0$ modes, and
 non-axisymmetric `modes', can.

Typically 64 grid cells were used per scale height in the $x$ and $y$
directions and slightly fewer in the $y$ direction. This fixes dissipation
above the grid scale. We also conducted runs with 128 grid cells, so as to
confirm convergence.

\subsubsection{Kink modes}

 First we exhibit some simulation results that reveal clearly the emergence of
the kink instability. For comparison with Section 2
we implement parameters which generate, at least initially, a `nearly
incompressible' run. 
The sound speed is set large with $c_s= 10\sqrt{5/3}$, and so
  $\beta=1.67\times 10^4$.
 The computational box is a `slab', with dimensions
 $4L\times 4L\times L$. The slab geometry is chosen so
 that 
 those growing kink modes which attack a two-stream channel 
 can fit into the horizontal domain. Finally, the initial condition is an
 order one amplitude two-stream MRI channel sprinkled with small amplitude
 perturbations. This choice eliminates muddy nonlinear interactions
 with other MRI modes and allows the channel to grow and collapse cleanly. 

\begin{figure}
\begin{center}
\scalebox{0.375}{\includegraphics{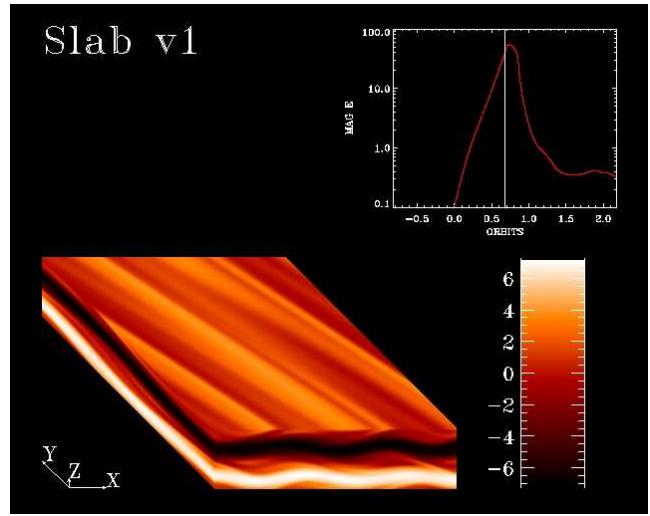}}
\caption{A linear kink mode attacking a two stream solution in
    slab geometry. The isosurfaces represent the $v_x$ component of the
    solution. In the top right is a graph of the time evolution of the
    magnetic energy scaled by thermal energy. 
   Note the large magnitude it can achieve before breakdown.
    Parameters are as stated in the text with $\beta=1.67\times 10^4$.}
\end{center}
\end{figure}

As expected, the two-stream channel grows exponentially to a point and then breaks
down. In accordance with the predictions of Section 2.4, the channel achieves
 large amplitudes before
disrupting, even when we increase the amplitude of the perturbations to
relatively large values: the channel's rapid growth helps it `outrun'
its parasites, initially.
 In Fig.~11 we present a coloured 3D plot of the isosurfaces of
$v_x$ near the peak of the channel growth. We have also inserted a time plot
of the total magnetic energy, divided by thermal energy, in the top right
corner.
 As is clear by
comparison with Fig.~2, a $k_z=0$ kink mode is attacking the two-stream channel,
forcing the characteristic sinusoidal pattern on each jet (a few streamers
radiating from each peak can also be seen, heralding the coming nonlinear
development). The mode
possesses $k_x/K= 1/2$ and $k_y/K=1/4$; or rather, $\theta_k$ is approximately
26 degrees and $k \approx 0.56$. The channel orientation $\theta$
 on the other hand, is near 45 degrees because it is nearly the fastest
 growing mode (cf.\ Section 2.2). So the kink mode wavevector is not
 perfectly aligned with the flow, but its amplitude is nearly optimal for
 maximum growth. In contrast, parasites that are
 perfectly aligned with the background flow (and fit into the slab) 
 possess wavenumbers closer to 0.35
 and 0.71, which yield growth rates of the same order or less.

\begin{figure}
\begin{center}
\scalebox{0.37}{\includegraphics{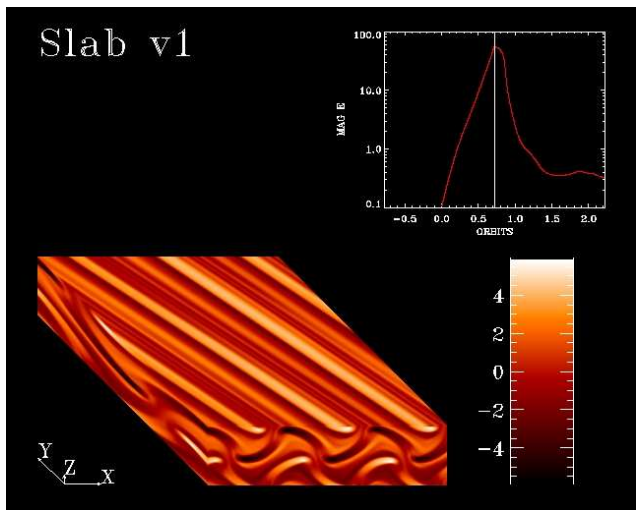}}
\caption{As before but at the nonlinear stage of the kink mode's
  evolution and just before the channel's catastrophic breakdown. Note how the
  billows have penetrated the adjacent jets.}
\end{center}
\end{figure}

The `billows' associated with the kink mode grow rapidly but never appear to
roll up into a series of `cats eye' vortices. Magnetic field is
swept up into each billow and magnetic tension exerts a torque against the
vortex' spin, as has been observed in other studies (Frank et al.~1996,
Dahlburg et al.~1997, Biskamp et al.~1998). Before secondary parasitic instabilities
destroy these structures 
they distort the two channels until they
interpenetrate each other, fragment, and then disintegrate catastrophically in
turbulence. The interpenetration stage, just before break down, is plotted in
Fig.~12.

\subsubsection{Compressible pinch modes}

 Next we isolate the compressible pinch mode in a simulation in the classic
`bar' geometry: $L\times 4L\times L$.  To aid entrance into this regime we
assign a lower sound speed $c_s=\sqrt{5/3}$ while keeping the Alfv\'enic speed
constant, and so $\beta=1.67\times 10^2$. These parameters are shared by the
remaining simulations presented in the paper.
 The initial
conditions are set to very low amplitude noise (the early stages of the system's
evolution will select the two-stream channel preferentially and so we need not seed
it explicitly).

As expected, the two-stream channel emerges cleanly from the early linear stage
of the system's evolution and continues to grow exponentially in the linear
regime. As the amplitude becomes larger we observe the channel
structure narrow to two thin planar jets accompanied by large spikes in
mass density. The magnetic pressure across each channel is sufficient to
efficiently squeeze matter onto the magnetic null surface. 
 This is revealed in Fig.~13, where the variation in density
between the troughs and the peaks is more than tenfold. This regime is reached
before the slow growing tearing modes of Section 2 (the only unstable modes
that can fit in the bar) achieve meaningful amplitudes. Recall that unstable
incompressible kink modes cannot fit into the bar geometry because the channel
has orientation
$\theta\approx\pi/4$ . 

\begin{figure}
\begin{center}
\scalebox{0.37}{\includegraphics{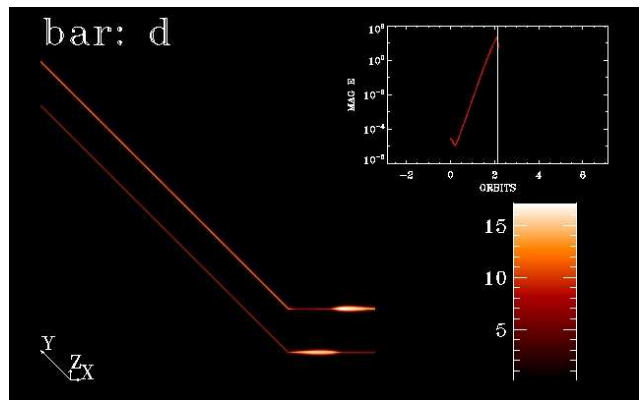}}
\caption{Two axisymmetric linear pinch modes on an extreme
    channel flow in bar geometry. In this plot the isocontours depict the total density $\rho$. The
    channel amplitude in this case is enormous, generating a magnetic
    pressure gradient sufficient to cause the massive variation in $\rho$
    between the centre of the jet and the evacuated region outside
    it. Here the $\beta$ is $1.67\times 10^2$.}
\end{center}
\end{figure}

Once the jets have
narrowed sufficiently they become susceptible to fast-growing \emph{compressible} parasitic
instabilities (Section 2.4.2) whose characteristic lengthscale is controlled by the jet width $l$,
not the overall channel wavelength $\lambda$. Consistent with Appendix A, the
dominant type of mode is the pinch instability, though on occasion it is
joined by a compressible kink mode.
 Typical perturbation profiles of the pinch mode,
 cf.\ Fig.~13, correspond relatively well to those examined in the
linear theory, cf. Fig.~10.
 The initial pinching motion subsequently generates a small magnetic island,
 or plasmoid, which is
convected along the jet. As a consequence, the plasmoid adopts the typical `droplet'
shape --- a leading blunt head and an attenuated tail (Biskamp 2000). Most
simulations show that
the plasmoids' growth continues into the nonlinear regime until the
perturbation impinges on its neighbouring jet or, as is more common, a fellow
plasmoid on the neighbouring jet, at which point the two flows collide and
break down in vigorous turbulence.
 A profile of a nonlinear
plasmoid can be observed in Fig.~16 in cube geometry. Those simulations which
exhibit the kink mode develop contorted jets more
prone to collision.

\subsubsection{Recurrent channels}

The two preceding runs dealt only with the initial stages of a system's
evolution, in which a clean two-stream solution emerged, reached large
amplitudes, and then collapsed due to a parasitic instability. We now turn our
attention to what
happens next. 

After the
inaugural outburst, the gas enters a disordered state, periodically organising
itself into a turbulent, two-stream structure and then relaxing
again. These aftershocks are much smaller than the first burst. 
 In particular, our results 
agree with the observations
of Bodo et al.\ (2008), confirming that computational boxes that are extended in the
$x$-dimension sustain channel bursts of smaller amplitude and frequency.
 In Figs 14 and 15
we plot channels at the tips of their energy peaks 
in a bar and a slab 
when
$\beta=1.67\times 10^2$. In
neither geometry does a two-stream channel achieve much coherence;
both suffer large amplitude perturbations even as they grow. But this is a
more pronounced
 in the slab case, where the channel really does
struggle to emerge from the turbulent background.

\begin{figure}
\begin{center}
\scalebox{0.37}{\includegraphics{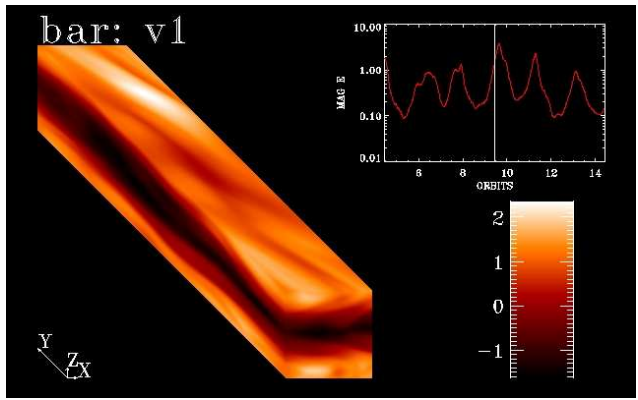}}
\caption{A $v_x$ snapshot of a channel peak in the later
    evolution of the MRI in the bar geometry.
  The channel is harrassed by large amplitude perturbations during its
    growth phase and never achieves the pure profiles of the previous cases;
    eventually it is folded up by the background disordered motions.
  Here the $\beta$ is $1.67\times 10^2$.}
\end{center}
\end{figure}

\begin{figure}
\begin{center}
\scalebox{0.37}{\includegraphics{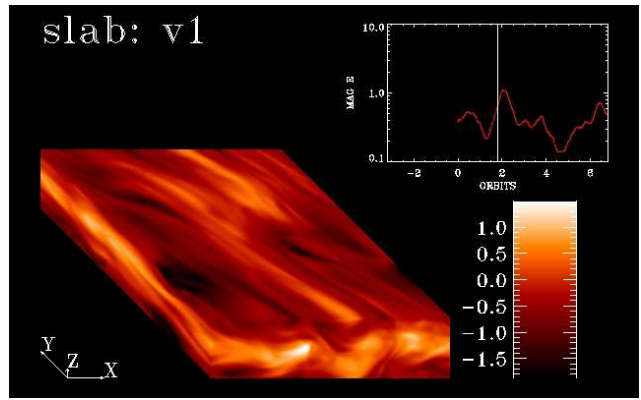}}
\caption{A $v_x$ snapshot of a channel peak in the later
    evolution of the MRI in the slab geometry.
The channel emerges with greater difficulty and the
perturbations upon it
are larger. It is more difficult to sustain coherence in the longer computational
domain when turbulence is present. Here the $\beta$ is $1.67\times 10^2$.}
\end{center}
\end{figure}

 Though both the channels in Figs 14 and 15 exhibit
`kinking motions' it
is incorrect to ascribe them
to parasitic modes (in the sense of
Section 2), as
is often done in the literature. Consistent low level turbulence is the
 salient feature of these runs, and its
 fluctuations are what ultimately destroy the channel.
In slab geometry more nonlinear interactions are available, because more
modes can fit into the box, and hence the
turbulence more developed. Channels in this case emerge with more
difficulty and are destroyed more quickly. (The same behaviour, 
for the same reasons, proceeds for greater $\beta$.) 
The large perturbations that the turbulence impresses upon the
channel structure are swept into
 the familiar
 Kelvin-Helmholtz-like pattern, but this is
accordance with the generic vortex dynamics of velocity shear, not the
particular dynamics of the linear kink mode. They are then folded-up and
thoroughly `mixed' by the turbulent eddies.

In contrast, when we choose parameters close to MRI marginal stability, as when $\beta$ is
lowered further, turbulent
fluctuations are absent in the troughs (there is only one growing MRI mode).
 This scenario permits the recurrent formation
and growth and destruction of very large-amplitude two-stream channels (cf.\
Sections 3.2.1 and 3.2.2), as 
each burst is of the same magnitude as the very first.
This eruptive behaviour has
been observed in the incompressible simulations of Lesur and Longaretti (2007).

\subsubsection{Cube geometry}

 Lastly, we look at the interesting case of MRI saturation in a
cube. The computational domain is set to $L\times L\times L$ and other
parameters take the fiducial values of the previous simulations with
$\beta=1.67\times 10^2$. 

In cubes two-stream flows are surprisingly robust and
 the basic channel structure
never breaks down completely. Initially, the gas evolves 
just as in the bar simulations:
 a channel grows to a large
amplitude and is subsequently attacked by a compressible pinch or kink mode.
 However, these instabilities
do not destroy the structure entirely, instead transforming it into 
 a disordered and fluctuating 
two-stream flow, characterised by plasmoid-type structures.
The key point is that these secondary structures never 
catastrophically impinge on each other or their
 neighbouring channels --- in fact, for most of the evolution the two streams
 appear relatively isolated from each other. The
 restrictive geometry of the cube does not offer sufficient room for the
 secondary structures to
deflect the two jets into each others' paths, and hence to destroy them.
One could say that the channels are `stiffened' by the short periodic boundaries.

\begin{figure}
\scalebox{0.37}{\includegraphics{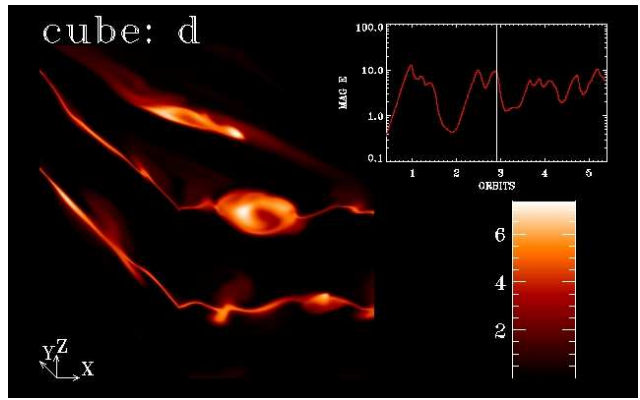}}
\caption{Saturated double jet in cube geometry as represented in
     density $\rho$. The upper jet
    exhibits a characteristic feature of the pinch mode's nonlinear
    saturation  --- a fat asymmetrical plasmoid. At future times the lower jet broadens
    and exhibits wake structures, the characteristic
    pattern of the saturated kink mode. Here $\beta=1.67\times 10^2$.}
\end{figure}

 Fig.~16 presents a snapshot from the saturated two-stream channel.
 Upon the upper jet
travels a  nonlinear plasmoid, 
looking much like the saturated endpoint of the
pinch mode in isolated magnetic jets. At later times the lower jet broadens out
and radiates `wakes', looking much like the saturated endpoint of the kink
mode in a magnetised jet (see Figs 4 and 5 in Biskamp et al.~1998). The appearance and
disappearance of these structures is intriguing, and perhaps suggest that the
system is oscillating about certain regions of phase space characteristic of
magnetic jets more generally (see Balbus and Lesaffre 2008 for more on
phase space analyses).

\section{Conclusion}

We briefly summarise the main results and ideas of the paper. 
Our principal goal was to understand in greater detail channel flows in the
nonlinear saturation of the MRI. We began by revisiting the
analysis of GX where we reinterpreted the two classes of parasitic mode that destroy
incompressible channels: they can be understood as a
`double kink mode' and a hybrid kink-pinch mode, analogues of classic
plasma instabilities. Resistivity was
added, which slightly increased the growth rates of the latter and introduced
slow growing `pinch-tearing modes', both changes a result of the availability
of field reconnection.

 An important observation made in
Section 2 is that, despite this family of parasitic instabilities, an
isolated small amplitude channel will inevitably grow to significantly large
amplitudes.
 A channel's exponential growth allows it to `outrun'
nascent parasitic modes until $b\gg 1$, at which point the parasites grow
sufficiently fast and overtake their host. 
Consequently, strong magnetic fields are inevitable whenever a
clean channel flow forms: destruction of these
flows is then usually due to those parasitic instabilities which can exploit
reconnection and so access the fund of energy stored in both the velocity and magnetic shear.
These observations are confirmed by numerical simulation in Sections 3.2.1 and
3.2.2. 

Next we investigate numerically the recurrent generation and destruction of channels
once the initial large amplitude channel is destroyed. The main observation we
make here is that channel destruction is qualitatively different from that
described by the linear parasite model. Such
channels form out of a persistent turbulent flow, whose fluctuations interfere
with its development, and ultimately tear it apart. The fact that channel
behaviour is weaker when $\beta$ and the radial box are bigger provides
evidence for this interpretation, because these changes enrich the nonlinear
turbulent dynamics by adding more active MRI modes.

These observations prompt the question:
 what role do channel solution play in real
accretion disks --- are they only artefacts of the shearing box's
assumptions? It is too early to answer this conclusively
 at this point. But a start can be
 made by incorporating a vertical lengthscale into the problem, such as would
 issue from vertical stratification.
The matter is complicated by the
 fact that in a stratified disk a linear MRI mode will not generally be a
  solution to the nonlinear equations. This suggests that isolated channels
 will never grow to the large amplitudes seen in Sections 3.2.1 and 3.2.2.
 Once $b\sim 1$, channels must begin interacting with other modes, and the
clean channel structure would break down. However, weak recurrent channel
 behaviour might remain a feature of MRI-induced turbulence in stratified
 disks, though highly resolved simulations in a semi-local shearing box may be
 required to reveal them.

\section*{Acknowledgments}
The authors would like to thank the anonymous referee for a careful review
which greatly improved the manuscript. This work was supported by a grant from
the Conseil R\'{e}gional de l'\^{I}le de France and a Chaire
d'Excellence awarded to S.~A.~B.\ by the French Ministry of 
Higher Education.

\appendix

\section{`Extreme channels' and their parasites}

This appendix offers a brief analysis of compressible channels at large amplitudes (`extreme
channels') and the
parasites which beset them. Throughout, $b\gg 1$ is maintained so that the
background shear, rotation, and imposed $B_z$ field are subdominant. We also
assume, as earlier, that the parasitic modes grow at a rate that far outstrips
the channel growth and the evolution of its $z$-profile. This permits us to
assume a steady background.

We present the main disruption mechanisms of intense
 compressible
channels and compare them with the incompressible case. 
 The qualitative
differences can be summarised neatly. When the MRI channels become
thinner and more intense the wavelengths of the unstable modes decrease
to significantly smaller values.
We also find that the modes change their
structure and generally localise on one or the other
jet. The principle modes are the compressible kink and pinch-tearing modes, and the
growth rates of the latter can approach the former as the plasma beta $\beta$
decreases to sufficiently low levels.

In the next subsection we outline the structure of the channel profiles adopted
and in the following sketch out the mathematical eigenproblem for the
compressible parasites. Once this is done the numerical results are summarised
and the main points demonstrated.

\subsection{Channel profile}

There is no straightforward analytical way to
obtain the profile of the compressible channel solution when $\beta$ is
small, in contrast to the incompressible case. In this regime, not only will
the solution grow exponentially, its $z$-profile will evolve with time as
magnetic pressure gradually squeezes the jet into a narrower and narrower
layer. Our analysis here is concerned with time-scales much shorter than this
evolution, and demonstrates general qualitative points of
behaviour.
 Therefore, we
take simple, but well-motivated, approximations.

We consider a compressible channel at a moment in time
when its amplitude dwarfs that of the imposed $B_z$ field and the background
Keplerian shear flow. For convenience, we rotate our $(x,y)$ frame so that $\ex$ points in
the direction of the channel flow $\v^\text{ch}$. The spatial profiles can
then be written as
\begin{align}
\v^\text{ch} &= v_0\,f(Kz)\,\ex\\
 \quad \B^\text{ch} &= B_0\,g(Kz)\,(\ex\,\cos\alpha+\ey\,\sin\alpha), \\
\rho^\text{ch} &= \rho_0\,h(Kz),
\end{align}
where $v_0$, $B_0$, and $\rho_0$ are dimensional constants, $f$, $g$, and $h$
are dimensionless functions, and $K$ is the vertical wavenumber of the
channel. 

The function $h$ can be related to $g$ if we assume, first, that the
gas is isothermal and, second, that to leading
order thermal pressure balances magnetic pressure in the $z$-component of the
momentum equation.
As to the second assumption, 
in a real simulation the total pressure possesses a small sink
centred on each channel which gradually attracts matter: thermal pressure
will increase to meet the large magnetic pressure gradients but can never exactly
cancel it. This pressure gradient should play a relatively minor role, but
may inhibit to some degree the
compressible kink mode

We set $P= c_s^2\,\rho$ where $c_s$ is the constant sound
speed; then the pressure balance implies
$$ h \approx 1+  \frac{1}{\beta}\,(1-g^2). $$
In so doing we take $\rho_0$ to be the density in the nearly evacuated regions
between the jets. The plasma beta is defined through $\beta\equiv 2 c_s^2/v_A^2$.

We employ the following analytic forms:
\begin{align}
f = \frac{\sinh [d\,\sin Kz ]}{\sinh d}, \qquad
g = \frac{\tanh [d\,\cos Kz]}{\tanh d}.
\end{align}
 These profiles
introduce the dimensionless parameter $d$ which is the
 ratio of the channel wavelength $\lambda=2\pi/K$ to
the width of the jet, $l$. Thus $d\equiv \lambda/l$. It
 quantifies the compression of the fluid achieved by
the magnetic pressure. It hence depends on $\beta$.
Roughly, the smaller $\beta$ the greater the compression and hence the greater
$d$. The profile choices above may not be immediately obvious but have the nice
property that they connect naturally to the incompressible channels of Section
2 in the double
limit $\beta,\,d^{-1}\gg 1$, on one hand, and the discontinuous current sheets, examined
by GX, in the limit $\beta,\,d^{-1}\ll 1$, on the other.
 What remains undetermined are the relationships between $v_0$, $K$,
$B_0$, which require solution of the nonlinear equations, something that can
be accomplished in each limit but is difficult in between.

\begin{figure}
\begin{center}
\scalebox{.5}{\includegraphics{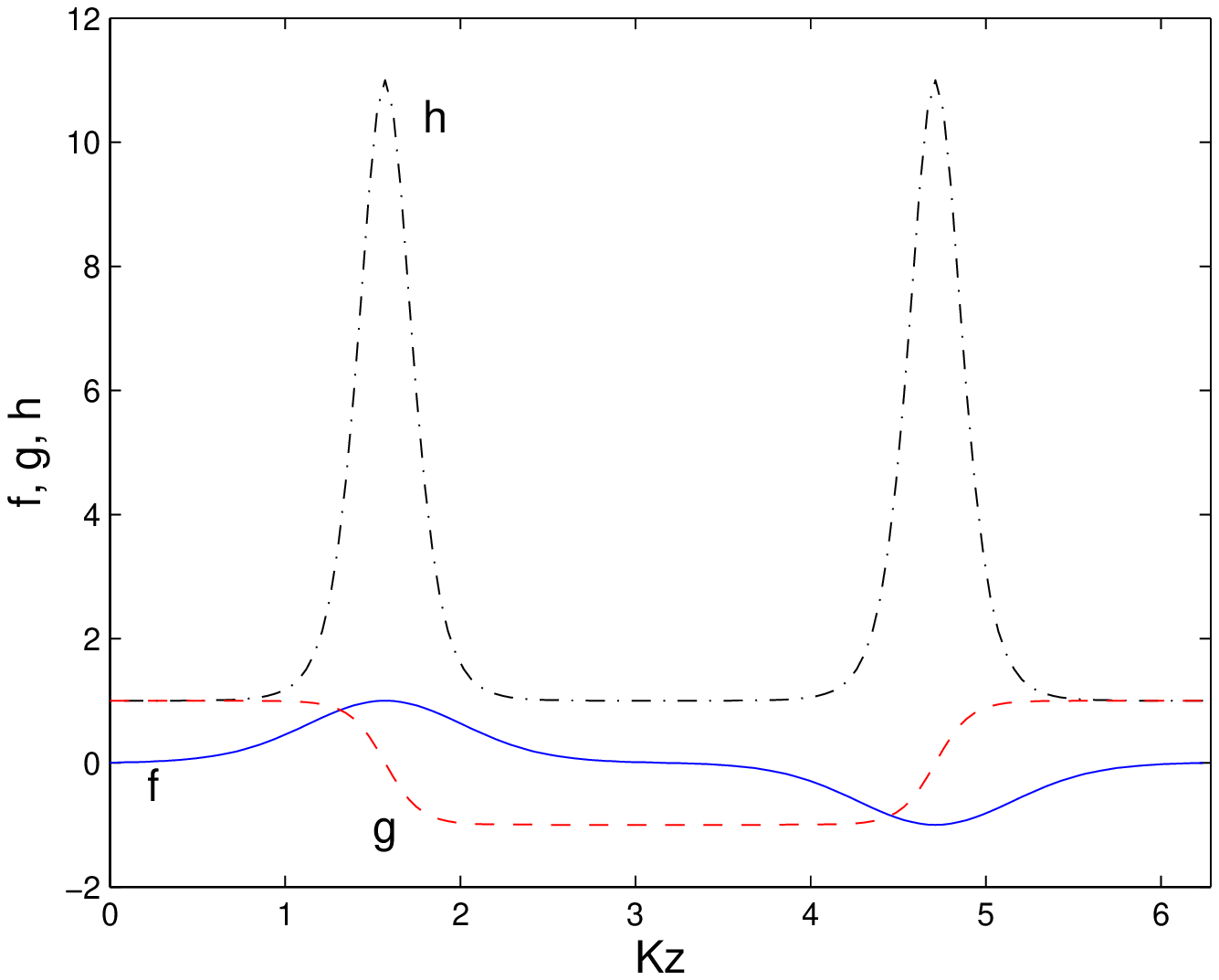}}
\caption{Plotted here are the dimensionless functions $f(Kz)$, $g(Kz)$, and
    $h(Kz)$ which characterise our approximation to the compressed channel solution. The
    parameters employed are $d=5$ and $\beta=0.1$. Note that the inward and
    outward jets are centred at $Kz= \pi/2$ and $3\pi/2$, altitudes which
    coincide with the magnetic null surfaces and the density maxima.}
\end{center}
\end{figure}

\subsection{Parasitic modes}

The approximate channel profiles introduced above are now perturbed by a small disturbance.
We assume that these parasitic modes possess a growth rate of order the channel amplitude
and thus we neglect the temporal evolution of the channel and the small $B_z$
and $v_z$ field associated with its gradual vertical compression. As earlier, we set
$$ \v=\v^\text{ch}+\v', \qquad \B=\B^\text{ch}+ \B', \qquad
\rho=\rho^\text{ch}+\rho',$$
linearise in the perturbations, and search for modes of the form $\propto
e^{ik_x x + ik_y y + \sigma t}$. Next we choose units so that $K=1$, $v_0
K=1$. The perturbations are scaled by $\rho_0$, $v_0$, and $B_0$. 
The governing compressible equations are now
\begin{align*}
&\sigma\v'= -i (\kk\cdot\v^0)\v' - \d_z\v^0\,v_z' 
-\frac{1}{M^2\,h}(i\kk+\ez\d_z)\rho'  \\
& \hskip0.9cm +\frac{1}{M_A^2\, h} \left\{i(\kk\cdot\B^0)\B' + \d_z\B^0\,B_z' -
  (i\kk+\ez\d_z)(\B^0\cdot\B')\right\} \\
& \sigma \B' = \left\{ i(\kk\cdot\B^0)\v'-i(\kk\cdot\v^0)\B'\right\} 
+\left\{\d_z\v^0\,B_z' - \d_z \B^0\,v_z' \right\} \\
& \hskip2.5cm -\left\{i(\kk\cdot\v') + \d_z v_z'  \right\}\,\B^0 -
\frac{1}{\Rm}\,(k^2-\d_z^2)\B' \\
& \sigma \rho'= -i(\kk\cdot\v^0)\rho' -(\d_z h)\,v_z' -
h\left\{i(\kk\cdot\v')+\d_z v_z'\right\},
\end{align*}
where most of the notation is carried over from Section 2.2 apart from the
Mach number which is defined through
$$ M= \frac{v_0}{c_s} $$
and the magnetic Reynolds number which we define differently
$$ \Rm= \frac{v_0}{\eta K}.$$
Note that $\beta= 2 M^2/ M_A^2$. Finally, the equilibrium fields are given by
$$ \v^0= f(z)\,\ex, \qquad \B^0= g(z)\,(\ex\,\cos\alpha+\ey\,\sin\alpha).$$
In the above, the incompressibility constraint has been replaced by the full
continuity equation: $\d_t\rho +\nabla\cdot(\rho\v)=0$.

As in Section 2, we have a Floquet boundary value problem and so assume the Floquet
ansatz and boundary conditions. This completes the statement of the
problem. The task now is to compute the growth rates $\sigma$ and associated
eigenfunctions given the following parameters: $M$, $M_A$, $\Rm$, $\alpha$, and $d$,
on one hand, and
$k_x$, $k_y$, and $k_z$, on the other.

\subsection{Numerical results}

We numerically solve the linear eigenvalue problem using a
Fourier pseudo-spectral method (Boyd 2000). This approach supplies us with an
approximation to the full spectrum for the given parameters. There are
unfortunately eight parameters, quite a number, 
and their various influences can obscure the main points we would like to
make. Therefore, only a few combinations are considered, those that are most
representative and which demonstrate the essential behaviour. 

First, we 
examine modes which appear in two-channel flows in a periodic box; in this domain there is sufficient
vertical space only for $k_z=0$ modes. Next we set $\alpha=-\pi/2$, so the $\B$
field is perpendicular to the $\v$ field. In practice, because of finite $\Rm$,
this angle will be slightly less but the effect we believe is small and
warrants neglect. Also we set $\Rm=500$, a representative value.
As for the Alfv\'{e}nic Mach number,
channel flows appear in MRI linear theory and in nonlinear simulations as
slightly sub-Alfv\'{e}nic, hence we set $M_A$ to values between 0.5 and 1.
 These choices still leave four parameters
parameters: $\beta$ (or $M$), $d$,  and $k_x$ and $k_y$. The latter two we replace by
magnitude and angle, $k$ and $\theta_k$.

\subsubsection{Compressible kink modes}

As in the incompressible case, the kink mode favours wavevectors oriented
along the background channel flow, $\theta_k=0$. When $d$ takes very small and
$\beta$ very large
values we obtain the same modes and behaviour as in Section 2, the only
difference being in the scaling. As we increase $d$ and decrease $\beta$ 
we enter the compressible
regime and the behaviour of the kink
mode changes in a number of ways.

 First, the thinner channels allow instability on a wider 
range of lengthscales, and the scale of the fastest
growing mode decreases. In the incompressible limit the fastest growing mode possesses
$k\approx 0.6$, while in the regime associated with $d=10$, $\beta=0.05$
 the fastest
growing mode possesses $k\approx 1.6$. 

Second, the structure of the kink
mode alters, 
particularly on the newly unstable shorter scales. In the incompressible case,
the kink mode attacked both channels concurrently. In the compressible case,
 higher $k$ kink modes localise on one or the other channel and thus appear in
 complex conjugate pairs (similar to the Type 2 and the pinch-tearing modes of Section
 2). However, kink modes of long horizontal wavelength cannot support this
vertical  localisation and at a critical (small) $k$ the two modes
 `detach' from each other, take different growth rates, and attack
  both channels concurrently.

 Lastly, the growth rate of the mode does not alter significantly. At smaller
 $\beta$ (and hence larger $M$) extra work must be expended in deforming the
 compressible plasma, but at smaller $\beta$ we take larger $d$ and so 
the jet is also thinner and
 exhibits a greater vertical shear, exacerbating instability. The two effects
 appear to roughly cancel.

\subsubsection{Compressible pinch-tearing modes}

The compressible 
pinch-tearing mode appears for all orientations of $\theta_k$. On decreasing $d$ it
transforms naturally into the $k_z=0$ incompressible pinch-tearing mode explored in Section
2.3.4. In contrast to the incompressible case, 
for general $d$ the compressible mode exists for orientations perpendicular to
$\B^\text{ch}$. It hence resembles the pinch modes exhibited in other studies (Wang
et al.~1988, Biskamp et al.~1998).

 The
structure of the mode is much the same as in Section 2, though the localisation is far more
pronounced when $d$ is large. This is seen in Figs 9 and 10  where we plot a
representative mode for $d=10$, $\beta=0.1$. There are two other striking changes as we approach
the compressible, large $d$ low $\beta$ limit.

 First, the mode grows
on a very large range of wavelengths, down to scales of order $(Kd)^{-1}$,
which reveals that it is strongly influenced by the jet width. As a
consequence, the fastest growing mode possesses a significantly shorter
wavelength than that of the overall channel structure $\lambda$ (and about half that of the
fastest growing kink mode).

 Second, the maximum growth rate is significantly
larger in the compressible regime. 
The pinch mode can  grow as  fast as the kink mode for a wide range of
parameters, in contrast to Section 2.
 The dominance of the pinch mode at
low $\beta$ (large $d$) has been reported
by Wang et al.~(1988) for an isolated current sheet, which is the case in our
analysis for certain parameters.
The rapid growth of the
pinch mode at small $\beta$ is due primarily to the thinness of the jets, a
configuration that greatly facilitates the reconnection of magnetic field:
 because oppositely directed field lines are physically so much closer it requires less
work to draw them together and break them. Hence, the increased growth rates
of the pinch instability reflect the ease with which the modes can access the fund of
energy stored in the background magnetic shear.

\end{document}